\documentclass[preprint,11pt,authoryear]{elsarticle}

\usepackage{amsmath, amssymb}
\usepackage{graphicx}
\usepackage{booktabs}
\usepackage{array}
\usepackage{microtype}
\usepackage[hidelinks]{hyperref}

\graphicspath{{figures/pdf/}{figures/}}
\newcommand{\Dthr}{D_{37}}
\newcommand{\Cthr}{C_{\mathrm{R5}}}
\newcommand{\EEthr}{EE_{\mathrm{thr}}}
\newcommand{\EEdet}{EE_{\mathrm{R5}}}
\newcommand{\rstar}{r^{*}}

\makeatletter
\def\ps@pprintTitle{\let\@oddhead\@empty\let\@evenhead\@empty
  \def\@oddfoot{}\let\@evenfoot\@oddfoot}
\makeatother

\journal{Icarus}

\begin{document}

\begin{frontmatter}

\title{Radiation limits on amino acid preservation and chiral measurement
at Oxia Planum}

\author{Dimitra Atri\corref{cor1}}
\cortext[cor1]{Corresponding author}
\ead{atri@nyu.edu}
\address{Space Exploration Laboratory, Center for Astrophysics and Space Science,
New York University Abu Dhabi, Abu Dhabi, United Arab Emirates}

\begin{abstract}
Amino acid enantiomeric excess (EE) is a potential biosignature, but its
measurement requires sufficient surviving material as well as preserved
chirality. We derive a criterion that separates abundance-limited from
racemization-limited preservation and apply it to the Mars Organic Molecule
Analyzer (MOMA) on the ExoMars Rosalind Franklin rover. The model combines cosmic-ray and radiogenic doses with laboratory
destruction kinetics and assumed ratios of interconversion to destruction rates. For an initially homochiral population of 1000~ppmw, the surviving EE is
58\% or 25\% when the total abundance reaches 1~ppmw for rate ratios of 0.04
or 0.10, respectively. This abundance criterion does not ensure that both
enantiomers can be quantified. Applying provisional estimates of 1~ppmw per
enantiomer and 10\% EE, serine at 2~m requires initial concentrations of
340 and 960~ppmw after cosmic-ray exposures of $10^{7}$ and $10^{8}$~yr,
respectively, under transferred fused-silica--perchlorate kinetics and a
1.57-MGy radiogenic dose. These concentrations exceed measured meteoritic serine abundances,
but do not predict the initial Oxia inventory. For organics retained in a closed host since deposition
3.9~Ga ago, the modeled radiogenic dose is 1.6--5.6~MGy across the adopted
compositions. In the lower-dose case, drilling to 2~m reduces the total dose
by only a factor of 1.5 after $10^{7}$~yr. The boundaries depend on mineral kinetics and organic residence time;
they distinguish preserved chirality from a measurable excess.
\end{abstract}


\end{frontmatter}

\section{Introduction}
\label{sec:intro}

The search for life on Mars depends on biosignatures that can survive the
ionizing radiation that destroys organic molecules in the martian near-surface.
Amino acids are a prime target, and their handedness is especially diagnostic:
terrestrial proteins contain almost exclusively L-amino acids, so an
enantiomeric excess (EE) in martian amino acids would be a potential
biosignature~\citep{Glavin2020}. However, an EE alone does not establish a biological
origin. Circularly polarized irradiation can produce a small primary imbalance in
amino acids~\citep{Modica2014,Glavin2020}, and subsequent chemical processing may
amplify it. Indigenous L-isovaline excesses in carbonaceous meteorites covary with
aqueous alteration~\citep{CroninPizzarello1997,GlavinDworkin2009,Glavin2021},
while large L-excesses in proteinogenic amino acids are difficult to separate from
terrestrial contamination (Section~\ref{sec:abiotic}).

The ExoMars \emph{Rosalind Franklin} rover will sample to 2~m depth at Oxia
Planum~\citep{Vago2017}, where the cosmic-ray dose is lower than at the surface. It carries the Mars Organic Molecule Analyzer (MOMA),
whose gas chromatography--mass spectrometry (GC--MS) mode can resolve
amino acid enantiomers after derivatization with
$N,N$-dimethylformamide dimethyl acetal (DMF-DMA)
~\citep{Goesmann2017}. Laser desorption mass spectrometry (LDMS) and other
GC--MS derivatization routes can guide the selection of compounds, but the chiral
measurement uses a separate DMF-DMA subsample. Laboratory tests constrain
derivatization-induced stereochemical change and column separation
~\citep{Boulesteix2024,Guzman2020}; recovery and peak integration in a martian
matrix remain compound-dependent~\citep{Boulesteix2023}.

MOMA requirement R5 specifies analysis of enantiomers at concentrations as low
as 1~ppmw (parts per million by weight)~\citep{Goesmann2017}. We apply it to the total analyte,
$C=[\mathrm{L}]+[\mathrm{D}]$. An estimate from the MOMA team (personal communication, 2026) puts reliable measurement at
${\sim}1$~ppmw for \emph{each} enantiomer and an EE of ${\gtrsim}10\%$. Below that EE,
variance and coelution could produce false positives. Flight-model tests were
still ongoing, so these are approximate values and not calibrated flight
limits. We adopt 1\% and 20\% as model EE reference values and compare the
10\% case with this MOMA-team estimate.

Galactic cosmic rays irradiate martian regolith to meter depths
~\citep{Atri2016,Atri2020,Atri2025,Pavlov2012}. Radiolytic destruction of amino acids is
considerably faster in some mineral mixtures than in pure solids~\citep{Pavlov2022,Pavlov2025}. Radioactive
decay of K, Th, and U adds a dose within the host rock that accumulates from
deposition and cannot be reduced by drilling~\citep{KminekBada2006}. Earlier
pure-solid kinetics implied modest loss from this channel; mineral-enhanced
destruction can change that conclusion. True L$\leftrightarrow$D
interconversion proceeds more slowly than destruction in the tested dry
solids~\citep{Bonner1979}, while an earlier aqueous period could change the
initial EE~\citep{BadaMcDonald1995}.

Laboratory studies have measured the destruction and the racemization of amino
acids by radiation~\citep{Bonner1979,Bonner1985,IglesiasGroth2011,Pavlov2022},
and have noted that radiolysis is much faster than radioracemization, so that
surviving amino acids may retain their chirality
\citep{Bonner1985,IglesiasGroth2011,Pavlov2022}. The contribution here is a quantitative separation of preservation from
measurement: we derive the crossover between abundance and EE loss, then
calculate whether the surviving major and minor enantiomers can meet specified
measurement criteria. We combine
both dose channels with destruction and racemization kinetics to
calculate the initial abundance required for a sampled horizon to reach the MOMA
requirement, the gain from drilling, and the EE of the surviving molecules. An
analytic criterion identifies which loss reaches its reference value first under
homogeneous, first-order kinetics. The dose model then gives the initial
abundance, depth, and exposure age needed to meet the assumed enantiomer
criteria. It also shows how radiogenic dose accumulated during burial reduces
the dose benefit of drilling and raises the required initial abundance.
Because compound- and matrix-specific interconversion rates are unavailable, we
vary their ratio to destruction. The estimates depend on the assumed transport
model and mineral matrix. Section~\ref{sec:methods} describes the dose model and
kinetics, Section~\ref{sec:results} gives the results, and
Section~\ref{sec:discussion} discusses target selection, the interpretation of a
detected EE, and the main limitations.

\section{Methods}
\label{sec:methods}

We propagate an initial amino acid population through radiolytic destruction and
radioracemization, then compare the surviving abundance with the R5 reference concentration and the surviving EE with specified reference values. A
depth-resolved galactic-cosmic-ray (GCR) dose model for Oxia Planum supplies the
radiation field. Published kinetic measurements supply the rate parameters and
the extrapolations described below.

\subsection{Radiation-transport model of the Oxia Planum subsurface}
\label{sec:dosemodel}

We use the Oxia Planum dose-depth output of the Geant4-based transport model
~\citep{Agostinelli2003,Allison2016,Atri2026}. It propagates primary GCR
protons, helium and heavier nuclei and their secondaries through 200 one-centimeter regolith
layers to 2~m depth. The hydrated-clay scenario has bulk density
2.2~g\,cm$^{-3}$, 5.5~wt\% water-equivalent hydrogen, and 1.0~wt\%
absorption-equivalent chlorine. The composition is assumed for hydrated clay-bearing material and has not been
measured at a specific Oxia outcrop. At
solar modulation $\varphi=660$~MV, the integral GCR flux at Oxia Planum is
$2.50$~cm$^{-2}$\,s$^{-1}$.

A silicon-slab calculation lies within 0.94--1.04 of the Mars Science Laboratory Radiation Assessment Detector (MSL/RAD)
detector-material reference~\citep{Hassler2014}. The agreement supports the
silicon surface normalization. The regolith dose and particle field at depth
remain unvalidated by this comparison.
Other martian subsurface transport calculations~\citep{Dartnell2007,Rostel2020}
use different compositions and dose definitions. \citet{Pavlov2022} computed an
independent Geant4 profile with protons and helium only, which gives a surface
dose rate about 30\% below the RAD value. Scaling our
dose rates by 0.5 approximately matches their reported isovaline survival at 2~m after 80~Myr (43\%
and 60\% for fused silica with and without perchlorate, against their 40\% and
60\%), so we use 0.5 as the low end of the dose-rate normalization in the
sensitivity analysis of Section~\ref{sec:team_sensitivity}. The agreement
constrains the combined dose--kinetics prediction; the depth profile itself
remains unvalidated by this survival comparison. We smooth the tabulated logarithmic dose
rate with a local cubic fit over a moving 21-cm window for display and
calculation. We retain the archived per-bin Monte Carlo standard errors
without reducing them by the number of smoothed bins, because the shared-history
covariance was not archived. These errors do not include kinetic or
compositional uncertainties.

The 200 layers extend from 0 to 200~cm. Their tabulated coordinates
mark the upper edges, so the deepest point represents 199--200~cm at the
nominal drill reach. A vacuum boundary immediately below it omits secondary
production in deeper rock and biases that bin low. To measure the bias, we
compared otherwise identical $10^{5}$-primary 2- and 6-m substrate runs and
applied their depth-dependent dose-rate ratio to the archived profile. The
fitted correction is 1.005 at 100--101~cm and 1.15 at 199--200~cm; the apparent
surface-to-deepest-layer attenuation falls from 15.6 to 13.5. A separate
6-m estimator gives 12.8, leaving about 5\% systematic uncertainty from the slab geometry at the
endpoint. The paired convergence runs used a newer Geant4 build than the
archived profile and differ in absolute normalization by about 10\%. Transferring
their ratio assumes that the relative boundary correction is less sensitive to
that change than the absolute dose.

Solar energetic particles add dose chiefly in
the upper centimeters~\citep{Pavlov2012}; omitting them makes the maximum
exposure ages of shallow samples upper limits and the required initial
abundances lower limits. For a sample held at depth $z$ over exposure age $t_{\mathrm{exp}}$,
\begin{equation}
D_{\mathrm{GCR}}(z,t_{\mathrm{exp}})
 = \dot D_{\mathrm{GCR}}(z)t_{\mathrm{exp}}.
\label{eq:cumdose}
\end{equation}
Burial and erosion histories require a time-dependent depth path.
A steadily exhuming horizon instead follows the dose-rate profile along its
path; because dose rises toward the surface, holding it at its present depth
throughout the interval overestimates its GCR dose. Martian mudstones with
measured cosmogenic exposure ages motivate the range of $t_{\mathrm{exp}}$
explored here~\citep{Farley2014}.

\subsection{Steady exhumation and the exposure-age parameter}
\label{sec:exhumation}

No cosmogenic exposure age has been measured at Oxia Planum, whereas erosion
rates for martian sedimentary rock are constrained by crater
statistics~\citep{KiteMayer2017}. It is therefore useful to express the same
calculation in terms of erosion rate. Substituting $z' = z + Et$ into
$D=\int_0^{\infty} \dot D[z(t)]\,dt$ for a horizon exhuming steadily at rate $E$
gives
\begin{equation}
D(z) = \frac{1}{E}\int_z^{\infty} \dot D(z')\,\mathrm{d}z'
     \;\longrightarrow\; \dot D(z)\,\frac{\lambda}{E},
\label{eq:exhum}
\end{equation}
where the right-hand limit holds for an exponential profile. The accumulated
dose is thus the depth integral of the dose-rate profile below the sample
divided by the erosion rate, and $\lambda/E$ plays the role of an effective
exposure age. We evaluate the integral form on the tabulated profile, with the
interior exponential supplying the tail below the deepest layer; this
gives an integral about 24\% larger than the closed form at the surface.
The continuation below 2~m is an extrapolation, not a measured deep-dose
profile; penetrating secondaries may give a different tail. The infinite-time
limit also assumes an erosion history long compared with $\lambda/E$. With
$\lambda = 70$~cm, erosion at $10$, $100$ and $1000$~nm\,yr$^{-1}$ corresponds
to effective exposure ages of $7.0\times10^{7}$, $7.0\times10^{6}$ and
$7.0\times10^{5}$~yr. A geomorphically active surface therefore accumulates far
less cosmic-ray dose than a $10^{8}$~yr constant-depth exposure would imply.

\subsection{Internal radiogenic dose}
\label{sec:radiogenic}

Decay of K, Th, and U also deposits energy in the host rock. Amino acids retained in an
Oxia sediment since deposition have also been irradiated by the decay of $^{40}$K, $^{232}$Th,
$^{238}$U and $^{235}$U in the host rock for the whole interval since
deposition. We approximate the buried host as a homogeneous infinite matrix:
$\beta$-particle transport is local on millimeter scales, whereas $\gamma$ rays
sample a larger surrounding volume. Within this approximation the radiogenic
dose is independent of sampling depth. It accumulates over the organic
residence time, taken here to equal the deposition age $t_{\mathrm{dep}}$.
This equality assumes emplacement at deposition, subsequent retention, and no
replenishment or late synthesis. Younger organics require a shorter integration
interval even in ancient sediment. Near a free surface, photon escape reduces
the local dose; the infinite-matrix value represents the prolonged buried
history rather than an exact present-day surface rate. For parent nuclides of mean life $\tau_i$
contributing present-day dose rates $\dot D_{i,0}$,
\begin{equation}
D_{\mathrm{rad}}(t_{\mathrm{dep}})
 = \sum_i \dot D_{i,0}\,\tau_i \left[\exp(t_{\mathrm{dep}}/\tau_i) - 1\right],
\label{eq:drad}
\end{equation}
where the exponential factors account for the higher parent abundances in the
past. The calculation assumes a closed radionuclide inventory and secular
equilibrium within the U and Th decay chains; redistribution and radon escape
are not modeled.
The total dose entering the kinetics of Sections~\ref{sec:destruction}
and~\ref{sec:racchannel} is then
\begin{equation}
D_{\mathrm{total}}(z, t_{\mathrm{exp}}, t_{\mathrm{dep}})
 = \dot D_{\mathrm{GCR}}(z)\, t_{\mathrm{exp}} + D_{\mathrm{rad}}(t_{\mathrm{dep}}).
\label{eq:dtotal}
\end{equation}

The abundances of K, Th and U in the Oxia Planum bedrock have not been measured,
so we treat the potassium content as the main free parameter. Thorium follows
from the mean martian K/Th of about 5300~\citep{Taylor2006} and uranium from
Th/U~$=3.8$. Applying the infinite-matrix conversion factors of
\citet{AdamiecAitken1998} and summing the $\beta$ and $\gamma$ contributions,
used here to approximate the dose to dispersed organics, we
adopt K~$=0.10$~wt\% as the baseline. This lies within the range of martian
basaltic meteorites (K~$=0.02$--$0.26$~wt\%; \citealp{Taylor2006}) and
reproduces the present-day dose rate of 130~$\mu$Gy\,yr$^{-1}$ computed by
\citet{KminekBada2006}. For a deposition age of 3.9~Ga appropriate to the Oxia
clay-bearing unit~\citep{QuantinNataf2021}, it gives $D_{\mathrm{rad}} = 1.57$~MGy.
Kminek and Bada also report 350~$\mu$Gy\,yr$^{-1}$ at 3~Ga and 0.74~MGy over the
last 3~Gyr; our baseline gives 0.89~MGy over 3~Gyr, because the decay of
$^{40}$K, which dominates, raises the dose rate in the past faster than the two
rates they quote imply.

To bracket the baseline, we use the Mars Odyssey Gamma Ray Spectrometer (GRS)
5$^{\circ}\times$5$^{\circ}$ element maps~\citep{Boynton2007, GRSdata} averaged over
bins within 5$^{\circ}$ (300~km, comparable to the instrument footprint) of
18.3$^{\circ}$N, 335.4$^{\circ}$E. This gives K~$=0.354\pm0.027$~wt\% and
Th~$=0.693\pm0.072$~ppm, both slightly above the global means of 0.319~wt\% and
0.589~ppm. Uranium is not measured by the instrument and is inferred as
U~$=0.182$~ppm from Th/U~$=3.8$. These abundances give 5.6~MGy at 3.9~Ga from
$\beta$ and $\gamma$ radiation, and 11.3~MGy if the $\alpha$ contribution is
added; the $\alpha$ dose is deposited within ${\sim}20~\mu$m of the U/Th-bearing
accessory phase, so this added contribution is not an estimate of the dose to the organic microphase. It is a
whole-rock average under complete energy absorption, not a strict upper bound
on the local dose to an organic molecule adjacent to a radioactive grain. The
present-day $\gamma$ rate in this case is 140~$\mu$Gy\,yr$^{-1}$.
Independent orbital-abundance modeling by \citet{GaryBicas2026} also finds
regional variation in the radiogenic environment; its global estimates do not
calibrate the dose to an organic-bearing Oxia microphase. Because the instrument senses only the upper few tens of centimeters
and averages over a footprint far larger than an outcrop, the GRS values describe
regional surface material, which we use as a high-K case for the bulk host rock.
Table~\ref{tab:drad} gives $D_{\mathrm{rad}}$ for a range of K contents and
deposition ages.

\begin{table}[htbp]
\centering
\caption{Radiogenic dose $D_{\mathrm{rad}}$ (MGy, $\beta+\gamma$) accumulated
since deposition for different potassium contents and
deposition ages. Th and U
follow from K/Th~$=5300$ and Th/U~$=3.8$. The baseline is K~$=0.10$~wt\%; the
regional GRS row instead uses its measured Th abundance (0.693~ppm)
and inferred U abundance (0.182~ppm), rather than the fixed K/Th ratio.
The rate is the present-day dose rate.}
\label{tab:drad}
\footnotesize
\begin{tabular}{r r r r r r}
\toprule
K (wt\%) & Th (ppm) & Rate ($\mu$Gy\,yr$^{-1}$) & 3.0~Ga & 3.5~Ga & 3.9~Ga \\
\midrule
0.05 & 0.09 & 65 & 0.45 & 0.61 & 0.78 \\
0.10 (baseline) & 0.19 & 129 & 0.89 & 1.23 & 1.57 \\
0.20 & 0.38 & 259 & 1.78 & 2.45 & 3.13 \\
0.354 (regional GRS) & 0.69 & 462 & 3.17 & 4.35 & 5.56 \\
0.50 & 0.94 & 647 & 4.45 & 6.13 & 7.83 \\
\bottomrule
\end{tabular}
\end{table}

Because Eqs.~\eqref{eq:destr} and~\eqref{eq:eedecay} depend on dose alone, the
radiogenic term enters both channels in the same way. It does not change the
regime criterion or the retained-excess identity derived in
Section~\ref{sec:regime}, both of which are dose-independent; it changes only
the depth- and age-dependent quantities.

\subsection{Radiolytic destruction channel}
\label{sec:destruction}

We represent solid-state amino acid radiolysis as first-order kinetics in
absorbed dose:
\begin{equation}
C(z, t_{\mathrm{exp}}, t_{\mathrm{dep}}) = C_0 \,
\exp\!\left[ -\,D_{\mathrm{total}}(z, t_{\mathrm{exp}}, t_{\mathrm{dep}})/\Dthr \right],
\label{eq:destr}
\end{equation}
where $\Dthr = 1/k_{\mathrm{radiolysis}}$ is the characteristic dose for survival
to $1/e$ and $C_0$ the initial concentration. $\Dthr$ values for each amino acid
(Table~\ref{tab:kinetics}) are taken from direct $\gamma$-irradiation
measurements: L-isovaline from \citet{Pavlov2022}, and the proteinogenic amino
acids from the solid-state $^{60}$Co constants of \citet{IglesiasGroth2011},
converted from their Solar-System dose normalization (14~MGy $\equiv$ 4.6~Gyr)
via $\Dthr[\mathrm{MGy}] = 30.43 / k_{\mathrm{dsc}}[10^{-10}\,\mathrm{yr}^{-1}]$.
For L-glutamic acid, $k_{\mathrm{dsc}}=1.60$, corresponding to
$\Dthr=19.0$~MGy. The Iglesias-Groth constants are inferred from
one 3.2-MGy endpoint using differential scanning calorimetry (DSC) enthalpy, with no
reported covariance. For valine and aspartic acid the inferred losses are only
1.9\% and 4.5\%, respectively, so their large $\Dthr$ values are particularly
sensitive to small endpoint errors.

An independent dry-solid $^{60}$Co experiment used four dose levels and high-performance liquid chromatography (HPLC)
quantitation~\citep{KminekBada2006}. It agrees for alanine
($\Dthr=8.87$ versus 8.55~MGy) but gives much smaller values for aspartic acid
(6.25 versus 69.5~MGy) and glutamic acid (5.80 versus 19.0~MGy). We retain both
datasets to test sensitivity to the kinetic source. Converting either laboratory dataset to the martian GCR
field further assumes that bulk absorbed dose is a sufficient chemical state
variable across radiation qualities and dose rates. Linear energy transfer (LET), track structure,
temperature, radical diffusion, and energy partition into organic-bearing
microphases are not represented by a scalar-dose conversion.
Because amino acids in the regolith are not pure solids, we scale the destruction
rate by the mineral-matrix enhancement factors measured by \citet{Pavlov2022} for
the L-isovaline series: $\times 2.1$ for a multi-amino acid (``cell-like'')
mixture, $\times 21$ for admixed fused silica, and $\times 35$ for fused silica
plus 2~wt\% dry magnesium perchlorate. The fused-silica+Mg-perchlorate case is an
empirical high-destruction case and has not been established for clay or
an Oxia endmember. These factors were
derived for L-isovaline, but \citet{Pavlov2022} also irradiated glycine, serine,
alanine, isovaline, $\alpha$-aminoisobutyric acid and 8-aminooctanoic acid
together in fused silica and found that fewer than 10\% of the amino acids
survived 2~MGy regardless of molecular mass. We therefore apply the same factors
to all amino acids as a baseline assumption. Valine, aspartic acid and glutamic
acid were not measured in a mineral matrix. A $\pm50\%$
compound-specific variation tests the effect of this transfer. As the single independent
anchor on this extrapolation, the Iglesias-Groth-derived L-serine $\Dthr$ combined with
the $\times 35$ factor predicts 9.7\% survival at 1~MGy, consistent with the measured
${\sim}10\%$ serine survival of \citet{Pavlov2022}. Independent measurements
found slower amino acid radiolysis in water ice than in Mars-like silicate
matrices~\citep{Pavlov2025}. Together, the Pavlov studies place the dry
fused-silica and fused-silica+Mg-perchlorate cases near the destructive end of
the measured laboratory range, while leaving the exact factor compound- and
matrix-dependent. The irradiations were done at room temperature, and glycine in
fused silica degraded less at $-50$ to $-55^{\circ}$C~\citep{Pavlov2022}, so
cooling could increase preservation. That result does not establish a
uniform temperature correction for all compounds and mineral mixtures. Glycine is excluded
from the chiral-detectability analysis because it is achiral.

\subsection{Radioracemization channel}
\label{sec:racchannel}

We define EE on the surviving molecules,
$EE=([\mathrm{L}]-[\mathrm{D}])/([\mathrm{L}]+[\mathrm{D}])$.
Within a homogeneous sample, equal destruction of L and D leaves EE
unchanged; symmetric L$\leftrightarrow$D interconversion reduces its
magnitude. A heterogeneous sample need not follow this rule in bulk:
differential survival of domains with different initial EE can reweight the
measured mixture without microscopic enantioselective destruction. The model
below represents a homogeneous sample.

For an interconversion rate $k_{\mathrm{truerac}}$ per unit dose in each direction
(L$\rightarrow$D and D$\rightarrow$L),
\begin{equation}
EE(z,t_{\mathrm{exp}},t_{\mathrm{dep}})=EE_0
\exp\!\left[-2rD_{\mathrm{total}}(z,t_{\mathrm{exp}},t_{\mathrm{dep}})/\Dthr\right],
\label{eq:eedecay}
\end{equation}
where $r=k_{\mathrm{truerac}}/k_{\mathrm{radiolysis}}$. This law preserves
the sign of $EE_0$; the criteria below use $|EE|$. Optical-activity decay
cannot isolate $r$ because it also records destruction of the parent
molecule.

Direct enantiomer measurements found 2--5\% reported racemization after
doses that destroyed 55--68\% of solid amino acids; aqueous sodium salts
showed 5--11\%, while hydrochloride salts showed little or none
~\citep{Bonner1979}. If the reported fractions are opposite-enantiomer
fractions among survivors, they correspond approximately to
$r=0.018$--0.066 in solids and 0.045--0.155 in aqueous salts. These
intervals pair extremes from different experiments and depend on the
reporting convention. We adopt $r=0.04$ and 0.10 as representative values for
dry solids and aqueous salts, respectively; neither is a measured Oxia rate.
Mineral-bearing leucine experiments found both channels accelerated
~\citep{Bonner1985}, with radiolysis constants 10 to 20 times larger than the
radioracemization constants, as quoted by \citet{Pavlov2022}. The conversion to our one-way interconversion ratio depends on the
published rate convention: a ratio defined using the EE-decay coefficient
would give $r=0.025$--0.05, whereas a one-way coefficient would give
$r=0.05$--0.1. We use this comparison as qualitative support for slower
interconversion, not as a direct calibration of $r$. No significant EE change
was detected for serine or isovaline in the irradiation conditions of
\citet{Pavlov2022}, although the D/L ratio of alanine rose from $0.06\pm0.03$ to
$0.20\pm0.05$ after 1~MGy, which they attributed to some radioracemization at a
rate well below the destruction rate. These results do not establish $r=0$, but
they are consistent with $r\ll1$. The optical-rotation ``$k_{\mathrm{rac}}$'' of
\citet{IglesiasGroth2011} is excluded because it conflates molecular loss
with L$\leftrightarrow$D interconversion.

Holding $r$ fixed across matrices assumes that a mineral enhances
interconversion and destruction equally, which has not been measured. A
planar $\alpha$-carbon radical formed by hydrogen abstraction can recover
either configuration or lose the parent molecule~\citep{Bonner1979}; other
reactions involving side chains, mineral radicals, or oxychlorines may change
the branching. The same uncertainty applies across amino acids. Despite lacking
an $\alpha$-hydrogen, isovaline underwent solid-state radioracemization in
\citet{BonnerIsovaline1979}, although \citet{Pavlov2022} did not detect
D-isovaline under its conditions. We therefore include isovaline in abundance
comparisons but do not assign it $r=0$.

L-serine is the baseline amino acid for the EE maps because it is proteinogenic,
so a biological initial excess is meaningful, and its matrix-scaled $\Dthr$
is consistent with one independent survival measurement. Isovaline is not used for the relation across amino acids discussed in
Section~\ref{sec:aspartic} because its meteoritic L-excess is associated with
parent-body aqueous alteration (Section~\ref{sec:abiotic}).

\subsection{Published R5 reference and MOMA-team per-enantiomer estimate}
\label{sec:criteria}

We set $\Cthr = 1$~ppmw from the published R5 concentration clause
(Section~\ref{sec:intro}) as a \emph{total-analyte} reference, with
$C=[\mathrm{L}]+[\mathrm{D}]$. The public numbered requirements and science
objective~O3 give no numerical EE criterion; O3 is qualitative (``analyze for the
presence and degree of ee in detected chiral organics''). The MOMA-team
estimate instead refers to approximately 1~ppmw \emph{per enantiomer} and
approximately 10\% EE. We keep the published reference and this estimate
separate.
Biological interpretation also depends on the identity of the amino acid and on corroborating
measurements. The MOMA
instrument paper notes that even an L-excess greater than 20\% in a
suite of amino acids would require additional evidence~\citep{Goesmann2017}.

To compare the two loss regimes, we use $\EEthr=1\%$ as a numerical reference
and also evaluate 10\% and 20\%. A drilled sample at
$(z,t_{\mathrm{exp}})$ meets both specified criteria when
\begin{equation}
C(z, t_{\mathrm{exp}}) \geq \Cthr\ (= 1~\mathrm{ppmw})
\quad\text{and}\quad
|EE(z, t_{\mathrm{exp}})| \geq \EEthr.
\label{eq:criteria}
\end{equation}
Each cell is classified as meeting both criteria, destruction-limited
(the abundance falls
below $\Cthr$ while the surviving material still exceeds $\EEthr$ in magnitude),
racemization-limited (the $|EE|$ falls below $\EEthr$ while abundance remains above
$\Cthr$), or fails both criteria. Section~\ref{sec:regime} also reports
$\EEdet$, the EE magnitude remaining exactly when abundance reaches $\Cthr$.
These classifications do not by themselves establish two detectable enantiomer
peaks.

For the MOMA-team estimate, both peaks must instead be present near or
above $C_{\mathrm{peak}}\simeq1$~ppmw and the excess near or above 10\%:
\begin{equation}
C_{\mathrm{minor}}=\tfrac12 C(1-|EE|)\gtrsim C_{\mathrm{peak}},
\qquad |EE|\gtrsim0.10.
\label{eq:teamcriteria}
\end{equation}
The minor peak controls the abundance condition. Neither inequality is a
calibrated completeness or false-positive boundary. If the minor peak falls below
$C_{\mathrm{peak}}$ while the major peak remains above it, the EE cannot be
measured, but a calibrated upper limit $U$ on its concentration and a measured major
concentration $M$ imply
\begin{equation}
 |EE| > \frac{M-U}{M+U},\qquad M>U.
 \label{eq:eebound}
\end{equation}
Here both concentrations must use the same sample basis and corrected analytical
response. We set $U=C_{\mathrm{peak}}$ only for the illustrative calculation;
a nominal sensitivity is not itself a statistical upper limit. The alternative
expression $|EE|>1-2U/C$ requires an independently known total concentration,
which a single detected peak does not supply.

\subsection{Regime analysis and parameter space}
\label{sec:regime}

Combining the relations above gives two limiting doses,
\begin{align}
D_{\mathrm{conc,max}} &= \Dthr \, \ln(C_0/\Cthr), \label{eq:dconc}\\
D_{\mathrm{EE,max}}   &= (\Dthr/2r) \, \ln(|EE_0|/\EEthr), \label{eq:dee}
\end{align}
These expressions assume $C_0>\Cthr$, $|EE_0|>\EEthr>0$, and $r>0$.
For $r=0$, EE is preserved; if either initial criterion fails, there is no
positive preservation interval under this model. The specified two-criterion
case is destruction-limited whenever
$D_{\mathrm{conc,max}} < D_{\mathrm{EE,max}}$, i.e.
\begin{equation}
r < \rstar = \frac{\ln(|EE_0|/\EEthr)}{2 \ln(C_0/\Cthr)}.
\label{eq:rstar}
\end{equation}
If $r$ is constant over the sample's radiation history, Eq.~\eqref{eq:rstar} is
independent of $\Dthr$ and of the dose profile. The \emph{regime} is then fixed
by $r$ and the two ratios $C_0/\Cthr$ and $|EE_0|/\EEthr$, while $\Dthr$
and $\dot{D}(z)$ set only \emph{where} in $(z, t_{\mathrm{exp}})$ the limit is
reached. For $\EEthr=1\%$ and $EE_0=100\%$, $\rstar$
ranges from 0.50 at $C_0=100$~ppmw to 0.25 at $C_0=10^{4}$~ppmw (0.33 at the
1000~ppmw baseline). Both values of $r$ lie below this boundary over the
explored parameter range. This holds for the 1\% criterion; stricter EE
criteria lower $\rstar$ (Fig.~\ref{fig:regime}).

Because the regime depends on $r$ once these two ratios are fixed,
we report the dimensionless distance
$g = \rstar/r$: the factor by which the assumed radioracemization-to-radiolysis
ratio would have to increase to change the outcome. Since $r$ has no empirical
distribution, $g$ is a ratio of rates and not a probability. For
$\EEthr=1\%$ and $EE_0=100\%$, $g=8.3$ for the dry-solid case and $3.3$
for the aqueous-salt case at $C_0=1000$~ppmw; at $C_0=10^{4}$~ppmw these
fall to $6.2$ and $2.5$, and at $C_0=10^{5}$~ppmw to $5.0$ and $2.0$.

For $r=0.10$, only a factor of two to three separates the case from the
crossover once $C_0$ is large. The classification is least certain for an
organic-rich sample whose radiation epoch overlapped persistent aqueous films,
and such material would benefit most from a dedicated determination of $r$.

Because $|EE_0|$ and $\EEthr$ enter Eq.~\eqref{eq:rstar} only through their ratio, increasing
the EE reference by a factor is equivalent to decreasing the assumed initial
excess by that factor. Solving $r = \rstar$ for the threshold gives the largest EE
criterion compatible with a destruction-limited outcome,
\begin{equation}
\EEdet = |EE_0| \, \exp\!\left[-2r \ln(C_0/\Cthr)\right]
       = |EE_0| \left(\Cthr/C_0\right)^{2r}.
\label{eq:eer5}
\end{equation}
Eliminating dose between the destruction and racemization
laws gives $|EE| = |EE_0| (C/C_0)^{2r}$, so $\EEdet$ is the excess
magnitude that a
population still retains at the instant its abundance falls to the reference concentration $\Cthr$. Within the constant-$r$ model it is fixed by $r$ and the
ratio $C_0/\Cthr$, independent of $\Dthr$, the dose profile, and any
assumption about analytical performance. In the general case,
\[
\ln(C_0/C)=\int k_{\mathrm{radiolysis}}\,dq,\qquad
\ln(|EE_0|/|EE|)=2\int k_{\mathrm{truerac}}\,dq,
\]
so a changing matrix, hydration state, radiation quality, or dose rate makes the
destruction-weighted effective $r$ path-dependent. Equation~\eqref{eq:eer5} then loses its
profile independence. Its numerical value also changes with $C_0$, $EE_0$, and
$r$.

At the baseline $C_0 = 1000$~ppmw with $EE_0 = 100\%$, $\EEdet = 57.5\%$ in dry solids ($r = 0.04$) and $25.1\%$ for the aqueous-salt case
($r = 0.10$); even for $C_0 = 10^{5}$~ppmw the latter value is $10.0\%$. For the
$EE_0=20\%$ case, the corresponding baseline values are 11.5\% and
5.0\%, and the aqueous-salt value
falls to 2.0\% at $C_0=10^{5}$~ppmw. An EE criterion yields a
destruction-limited classification only when it lies below the corresponding
$\EEdet$. At the baseline $EE_0=100\%$ and $C_0=1000$~ppmw, even a 20\%
criterion remains destruction-limited for both $r$ values, whereas the
aqueous-salt case crosses into EE-limited behavior above
$C_0\simeq3.1\times10^{3}$~ppmw.

Quantification of both enantiomer peaks requires a peak-specific criterion.
Under the total-analyte convention, the two concentrations are
$C_{\mathrm{major,minor}}=(1\pm|EE|)C/2$. At $C=1$~ppmw, the baseline
$|EE|=57.5\%$ case therefore contains 0.7875 and 0.2125~ppmw, while the
$|EE|=25.1\%$ case contains 0.6255 and 0.3745~ppmw. These are modeled enantiomer
abundances. Under the ${\sim}1$~ppmw-per-enantiomer estimate,
$C\gtrsim2C_{\mathrm{peak}}/(1-|EE|)$, requiring total concentrations of
approximately 4.7 and 2.7~ppmw at 57.5\% and 25.1\% EE, respectively. Laboratory method limits have compound- and test-specific definitions
and are not interchangeable with this estimate~\citep{Boulesteix2024}.

We explore the depth range $z \in [0, 199]$~cm, exposure age
$t_{\mathrm{exp}} \in [10^{5}, 10^{9}]$~yr, six amino acids spanning the measured
radiolytic resistance, the four matrix scalings defined in Section~\ref{sec:destruction}, the dry-solid
and aqueous-salt racemization sensitivities, and $EE_0 = 100\%$ (with
$EE_0 = 20\%$ as
a sensitivity case). The baseline initial concentration is $C_0 = 1000$~ppmw,
varied over $10$--$10^{5}$~ppmw in the regime analysis.

\subsection{Earlier aqueous history and sedimentary racemization}
\label{sec:aqmethod}

The two radiation channels above describe the post-burial constant-depth epoch.
To show how an earlier aqueous period could alter the initial EE, we add a
separate sedimentary racemization sensitivity calculation:
$EE/EE_0=\exp(-2 k_{\mathrm{rac}}(T)\,\tau_{\mathrm{aq}})$ with an Arrhenius
$k_{\mathrm{rac}}(T)=A\exp(-E_a/RT)$ ($E_a\approx93$--$110$~kJ\,mol$^{-1}$),
$\tau_{\mathrm{aq}}$ the duration of the aqueous period, and constants from \citet{Steen2013}, as corrected
by \citet{SteenCorrection2015}, following
\citet{BadaSchroeder1972}. These rate constants describe bulk marine sedimentary
organic matter,
extrapolated from 49.5--105$\,^{\circ}$C, with aspartic acid (Asp) pooled as Asx
(Asp+Asn) and glutamic acid (Glu) as Glx (Glu+Gln). The calculation gives an order-of-magnitude
estimate and omits pH, salinity, mineral catalysis, coupled aqueous destruction,
and time-dependent temperature.
Isovaline, lacking an $\alpha$-hydrogen, does not racemize by this mechanism (the
basis of its role as a meteoritic marker; \citealp{GlavinDworkin2009}), so
$k_{\mathrm{rac}}=0$ for it. Under sequential, uncoupled epochs the excess is
$EE_0\exp(-2k_{\mathrm{rac}}\tau_{\mathrm{aq}})\exp(-2rD/\Dthr)$, evaluated over
$\tau_{\mathrm{aq}}=10^{3}$--$10^{7}$~yr at 273~K.

\begin{table}[htbp]
\centering
\caption{Characteristic dose $\Dthr$ (the dose for survival to $1/e$) for amino acid destruction in pure solid
and after the L-isovaline-derived matrix enhancements measured by
\citet{Pavlov2022} ($\times21$ for fused silica; $\times35$ with
Mg-perchlorate). Pure-solid L-isovaline is from that study; proteinogenic
values come from the single-endpoint DSC results of
\citet{IglesiasGroth2011}. The matrix factors are transferred across amino acids.
Independent HPLC kinetics reverse the order of aspartic acid and alanine
(Section~\ref{sec:destruction}), and the valine value rests on a
1.9\% DSC decrement.}
\label{tab:kinetics}
\footnotesize
\setlength{\tabcolsep}{3.5pt}
\begin{tabular}{l c r r r >{\raggedright\arraybackslash}p{2.8cm}}
\toprule
Amino acid & $\alpha$-H & \begin{tabular}{@{}c@{}}$\Dthr$ pure\\(MGy)\end{tabular}
 & \begin{tabular}{@{}c@{}}$\Dthr$ +silica\\(Gy)\end{tabular}
 & \begin{tabular}{@{}c@{}}$\Dthr$ +silica\\+Mg(ClO$_4)_2$\\(Gy)\end{tabular} & Notes \\
\midrule
L-isovaline     & no  & 10.1 & $4.8\times10^{5}$ & $2.9\times10^{5}$ & non-proteinogenic $\alpha$-methyl amino acid \\
L-aspartic acid & yes & 69.5 & $3.3\times10^{6}$ & $2.0\times10^{6}$ & value depends on kinetic source \\
L-glutamic acid & yes & 19.0 & $9.1\times10^{5}$ & $5.4\times10^{5}$ & value depends on kinetic source \\
L-alanine       & yes & 8.55 & $4.1\times10^{5}$ & $2.4\times10^{5}$ & proteinogenic \\
L-serine        & yes & 15.0 & $7.1\times10^{5}$ & $4.3\times10^{5}$ & proteinogenic; consistent with one survival endpoint \\
L-valine        & yes & 166  & $7.9\times10^{6}$ & $4.8\times10^{6}$ & negligible DSC signal change \\
\bottomrule
\end{tabular}
\end{table}

\begin{figure}[htbp]
\centering
\includegraphics[width=\textwidth]{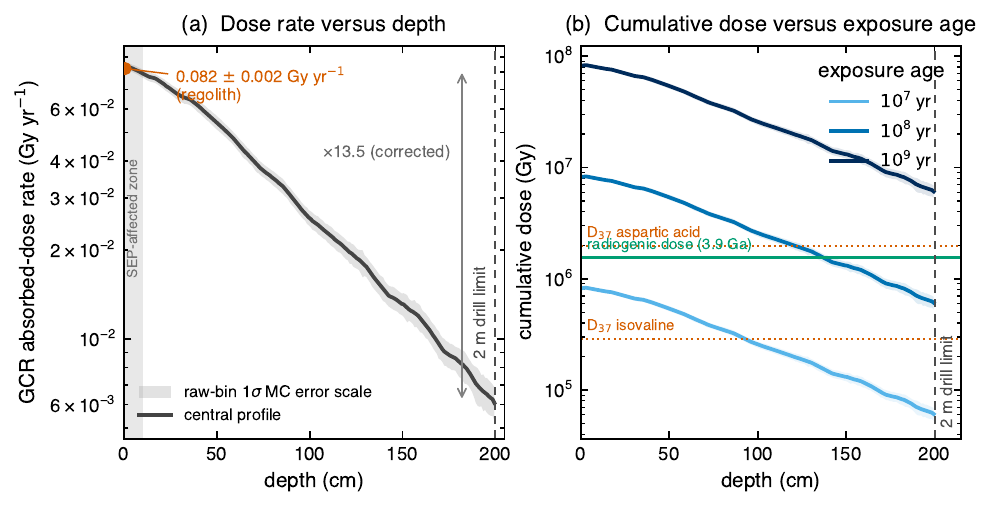}
\caption{Radiation dose in the Oxia Planum subsurface. (a) GCR absorbed-dose
rate with per-bin Monte~Carlo 1-$\sigma$ errors around the smoothed
profile. The finite-slab correction measured from the convergence runs is 1.005 at 100--101~cm and 1.15 at 199--200~cm;
the corrected surface-to-deepest-layer attenuation is 13.5 with about 5\%
residual geometric uncertainty. Depths mark the midpoints of 1-cm layers.
The gray strip marks the upper 10~cm, where solar energetic particles, which
are not modeled, add dose. (b) Cumulative GCR dose for three constant-depth
exposure ages, compared with the 1.57-MGy radiogenic dose and $\Dthr$ for
L-isovaline and L-aspartic acid in the fused-silica+Mg-perchlorate case.}
\label{fig:dose}
\end{figure}

\section{Results}
\label{sec:results}

\subsection{Dose accumulation in the drill-accessible subsurface}
\label{sec:doseatten}

The boundary-corrected dose profile declines from a surface absorbed-dose
rate of $0.0822\pm0.0022$~Gy\,yr$^{-1}$ to
$(2.55\pm0.16)\times10^{-2}$~Gy\,yr$^{-1}$ in the 100--101~cm layer, a factor
of 3.23. It reaches $(6.09\pm0.79)\times10^{-3}$~Gy\,yr$^{-1}$
in the deepest 199--200~cm layer, a factor of 13.5 below the surface
(Fig.~\ref{fig:dose}). Because cumulative dose scales linearly with cosmic-ray
exposure age, the dose received by an amino acid depends on both the depth of
the sample and the time its present horizon has spent within roughly a
meter of the surface. The dose falls progressively with depth, with a
${\sim}15\%$ boundary bias at the deepest bin (Section~\ref{sec:dosemodel}).
A surface outcrop accumulates 1~MGy in approximately $1.2\times10^{7}$~yr at
the modeled GCR dose rate.

The depth-independent channel is of comparable size. The baseline radiogenic
dose of 1.57~MGy (Section~\ref{sec:radiogenic}) equals the GCR dose accumulated
in $1.9\times10^{7}$~yr at the surface, $6.2\times10^{7}$~yr in the
100--101~cm layer, and $2.6\times10^{8}$~yr in the 199--200~cm layer. For shorter exposure ages the
radiogenic term dominates the total dose, and drilling does not reduce it.
Under steady exhumation (Eq.~\ref{eq:exhum}) at $100$~nm\,yr$^{-1}$ the surface
GCR dose is $7.2\times10^{5}$~Gy, below the radiogenic term at every depth; at
$10$~nm\,yr$^{-1}$ it is $7.2\times10^{6}$~Gy and the cosmic-ray channel
dominates in the upper meter.

\subsection{Separation of the two loss channels in the baseline calculation}

\begin{figure}[htbp]
\centering
\includegraphics[width=\textwidth]{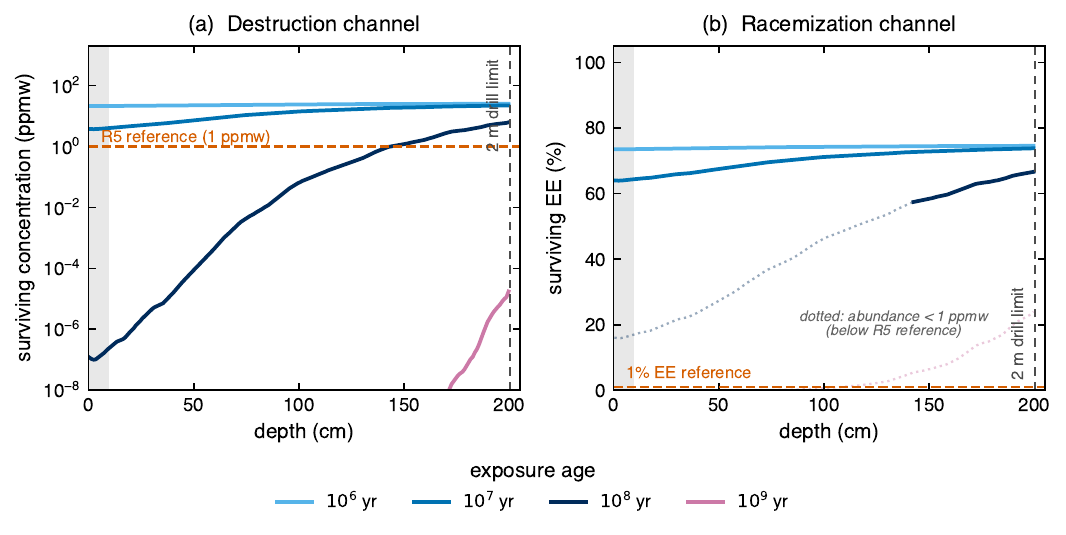}
\caption{Surviving (a)~concentration and (b)~enantiomeric excess versus depth for
L-serine in the fused-silica+Mg-perchlorate case ($\Dthr=4.3\times10^{5}$~Gy, $r=0.04$) at exposure ages of $10^{6}$,
$10^{7}$, $10^{8}$ and $10^{9}$~yr. The gray strip marks the upper 10~cm, where
solar energetic particles are not modeled. Both panels use the total
absorbed dose: depth-dependent GCR exposure plus the adopted depth-independent
radiogenic contribution accumulated over the Oxia deposition age. The dashed line in panel~(a)
is the 1~ppmw total-analyte R5 reference and the dashed
line in panel~(b) is the 1\% EE reference. In panel~(b) the EE curve is
drawn dotted wherever the corresponding abundance in panel~(a) has fallen below
1~ppmw. Values are plotted at the midpoints of the 1-cm layers; the deepest
point at 199.5~cm represents the 199--200~cm layer and therefore the full 2-m
modeled interval.}
\label{fig:survival}
\end{figure}

Abundance and EE reach their specified reference values at different doses
(Fig.~\ref{fig:survival}). For the baseline scenario
(L-serine in the fused-silica+Mg-perchlorate case,
$\Dthr = 4.3\times10^{5}$~Gy, dry-subsurface $r = 0.04$), the concentration falls
to the 1~ppmw R5 reference at
$D_{\mathrm{conc,max}} = 3.0\times10^{6}$~Gy. The enantiomeric excess reaches the
1\% reference at
$D_{\mathrm{EE,max}} = 2.5\times10^{7}$~Gy. From Eqs.~\eqref{eq:dconc} and~\eqref{eq:dee}, their ratio is
$D_{\mathrm{EE,max}}/D_{\mathrm{conc,max}} = \rstar/r = g$, the same margin
introduced in Section~\ref{sec:regime}, here equal to 8.3.
At the R5 reference concentration the surviving excess magnitude is
$\EEdet = |EE_0|(\Cthr/C_0)^{2r}$ (Eq.~\eqref{eq:eer5}), independent of $\Dthr$ and dose
profile at fixed $r$: by the
time radiolysis has reduced an initially homochiral ($EE_0 = 100\%$) 1000~ppmw
population to the 1~ppmw R5 reference, it retains ${\sim}58\%$ EE in the dry case
($r = 0.04$) and ${\sim}25\%$ for the aqueous-salt case ($r = 0.10$),
both above the 1\% reference (Fig.~\ref{fig:survival}b). Meeting the R5 requirement additionally requires the abundance in Fig.~\ref{fig:survival}a to
remain at or above 1~ppmw.

After $10^{9}$~yr, in the deepest modeled
layer (199--200~cm, corresponding to the full 2-m drill reach), the GCR-only
calculation gives $C=6.8\times10^{-4}$~ppmw and an EE of 32.1\% for the
surviving molecules.
Including the adopted radiogenic dose reduces these values to
$C=1.7\times10^{-5}$~ppmw and 23.9\% EE, respectively. Figure~\ref{fig:survival}
shows the total-dose case. Its concentration is below 1~ppmw at every modeled
depth, so the entire corresponding EE curve is drawn dotted. The modeled EE
therefore remains nonzero even when too little material is left to measure.

\subsection{R5-referenced preservation map}

\begin{figure}[htbp]
\centering
\includegraphics[width=\textwidth]{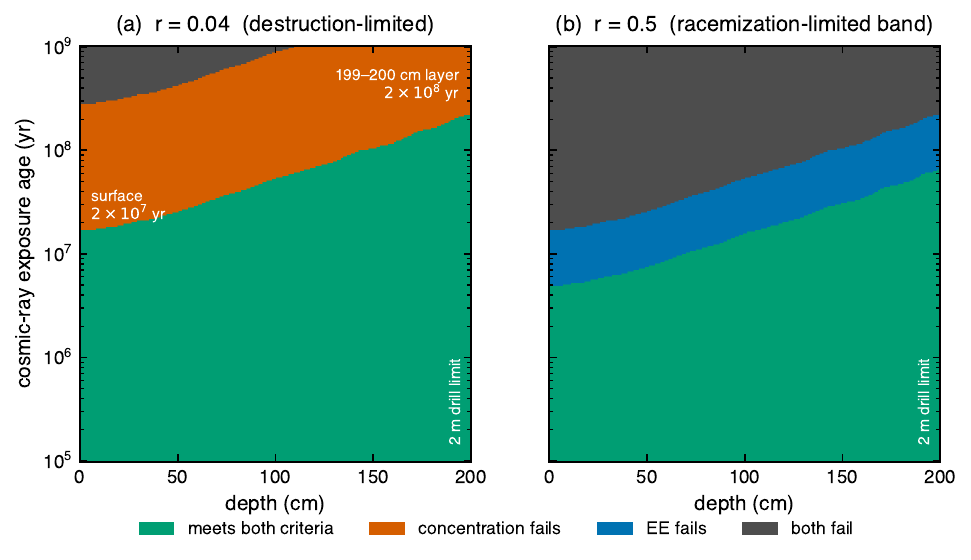}
\caption{R5-referenced preservation map for L-serine in the fused-silica+Mg-perchlorate
laboratory case at
$C_0=1000$~ppmw, $EE_0=100\%$, and the 1\% EE reference. L-serine is
the baseline amino acid because it is proteinogenic, so a biological $EE_0$ is
meaningful for it, and because its transferred matrix-scaled $\Dthr$ is consistent
with an independent measured survival fraction.
Green marks cells that meet both criteria, orange cells where only the
concentration criterion fails (destruction-limited), blue cells where only the
EE criterion fails (racemization-limited) and gray cells where both fail.
(a)~Dry-solid case, $r=0.04$: the boundary is destruction-limited; the labels
give the maximum exposure ages at the surface and in the 199--200~cm layer.
(b)~Control with $r=0.50$ (above $\rstar=0.33$): a racemization-limited band
appears.}
\label{fig:map}
\end{figure}

Classifying every (depth, exposure age) cell against the R5 reference concentration and 1\% EE reference yields the preservation map
(Fig.~\ref{fig:map}). For the baseline amino acid and
matrix, the region that meets both criteria lies below a single boundary
that rises from a maximum exposure age of
$1.69\times10^{7}$~yr at the surface to
$5.46\times10^{7}$~yr in the 100--101~cm layer and $2.29\times10^{8}$~yr in the
199--200~cm layer.

The radiogenic channel consumes 53\% of the limiting dose $D_{\mathrm{conc,max}}$ of this amino acid before any cosmic-ray exposure accumulates, so the corresponding GCR-only
maximum exposure ages are higher: $3.60\times10^{7}$, $1.16\times10^{8}$ and
$4.9\times10^{8}$~yr. Because the radiogenic term is depth-independent it
subtracts the same amount from the numerator at every depth, so the
\emph{ratio} of the maximum exposure ages at two depths is unchanged at 3.23 between the
surface and the 100--101~cm layer; what changes is the absolute exposure age each depth allows. When the concentration criterion controls, these ages are
$[\Dthr\ln(C_0/\Cthr)-D_{\mathrm{rad}}]/\dot D_{\mathrm{GCR}}(z)$
and depend logarithmically on the assumed initial abundance
($C_0 = 1000$~ppmw here). The limiting channel is set by $r$ relative to the
crossover $\rstar(C_0)$ (Eq.~\eqref{eq:rstar}), which is independent of $\Dthr$ and of the dose
profile. For the
1\% reference, baseline $C_0$, and $r=0.04$, the entire
boundary is abundance-limited by destruction. At fixed $r$, $C_0$ and $EE_0$,
depth shifts the age boundary while the limiting channel remains unchanged.
Raising $r$ to 0.50, above the
crossover $\rstar = 0.33$, produces a racemization-limited band at the modeled
preservation boundary (Fig.~\ref{fig:map}b); this value serves only as a
control.

\subsection{Sensitivity to kinetic parameters and the EE criterion}

\begin{figure}[htbp]
\centering
\includegraphics[width=0.82\textwidth]{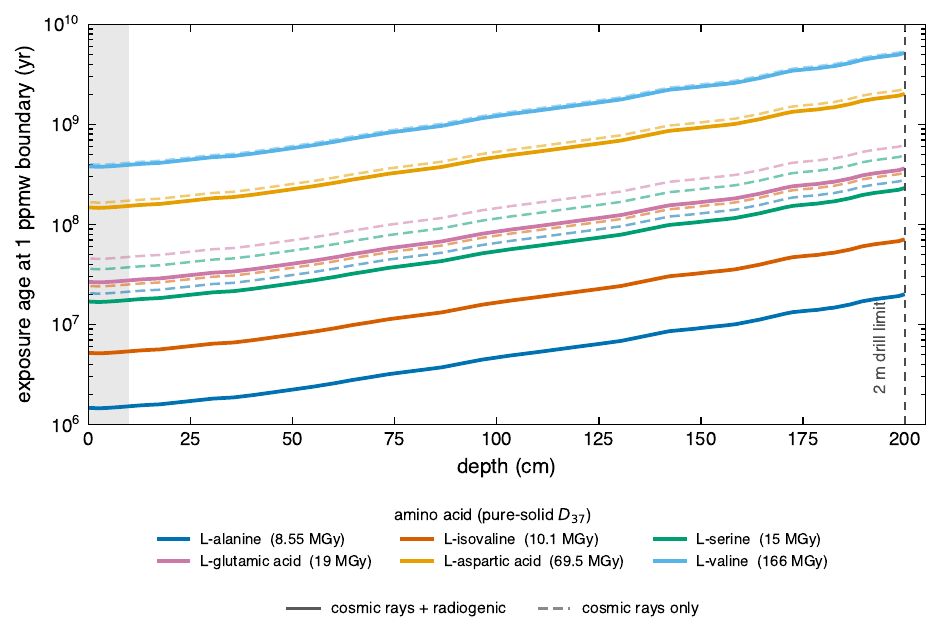}
\caption{Maximum constant-depth exposure age at which the concentration remains above
1~ppmw for six amino acids ($C_0=1000$~ppmw; fused silica+Mg-perchlorate; cosmic rays plus the 1.57-MGy
radiogenic dose for solid curves and cosmic rays only for dashed curves), ordered from
radiation-sensitive to radiation-resistant using the baseline kinetic
compilation (Table~\ref{tab:kinetics}). The
independent Kminek and Bada constants reverse part of this ordering. The deepest
endpoints represent the 199--200~cm layer and use the boundary-corrected dose
rate. The gray strip marks the upper 10~cm, where solar energetic particles are
not modeled.}
\label{fig:ceiling}
\end{figure}

At fixed matrix, the compound-specific $\Dthr$ determines the position of the
R5-referenced preservation boundary (Fig.~\ref{fig:ceiling}). Across the six amino acids of
Table~\ref{tab:kinetics} in the fused-silica+Mg-perchlorate case at the baseline
$C_0=1000$~ppmw, the maximum exposure age at the surface spans a factor of about 260 in the Iglesias-Groth baseline,
from $1.5\times10^{6}$~yr for L-alanine to $3.8\times10^{8}$~yr for L-valine,
with L-aspartic acid ($1.5\times10^{8}$~yr) and L-glutamic acid
($2.7\times10^{7}$~yr) intermediate. The spread is wider than the factor of 20
obtained from the cosmic-ray channel alone, because the depth-independent dose
subtracts a fixed dose from the limiting dose of every amino acid and so weighs most heavily
on the amino acid with the smallest limiting dose. These values follow the Iglesias-Groth
kinetic dataset. The Kminek--Bada HPLC constants give a
maximum surface exposure age of 2.2~Myr for alanine. For aspartic and glutamic
acid the limiting doses at $C_0=1000$~ppmw (1.2 and 1.1~MGy) are below the
radiogenic dose, so no exposure age is permitted at any depth, which reverses
the order of aspartic acid and alanine. For the present dose profile, each baseline curve is raised
by a factor of 3.23 in the 100--101~cm layer. The matrix factors multiply destruction rates by 21 and 35 relative to
pure solids. Maximum exposure ages do not scale by the same factors after
subtracting the radiogenic dose; for a sufficiently small dose budget, a
matrix change removes the permitted exposure interval altogether. If the L-isovaline-derived matrix
factor is varied independently by $\pm50\%$ for each amino acid, the order of adjacent amino acids can change.

\begin{figure}[htbp]
\centering
\includegraphics[width=\textwidth]{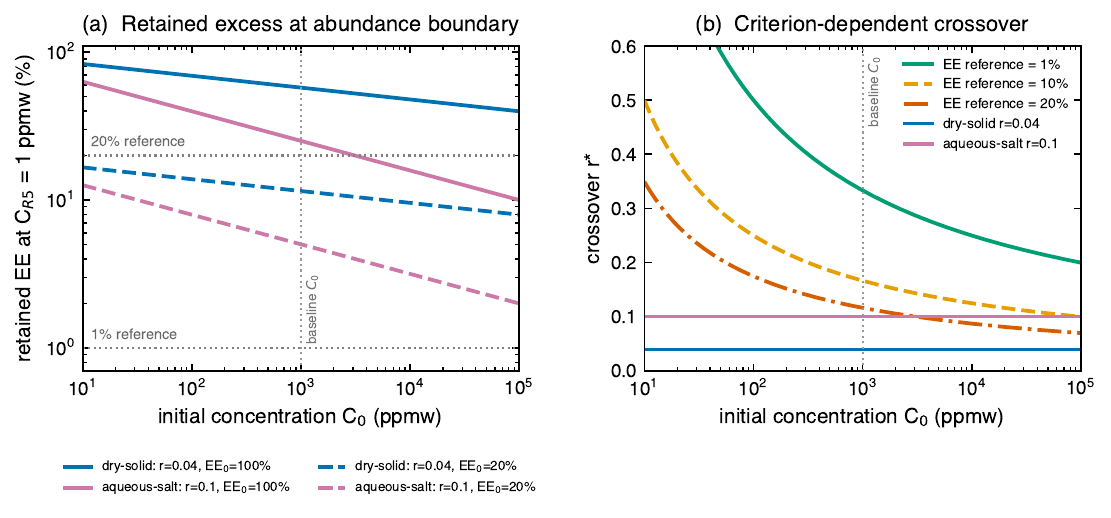}
\caption{Dependence of the regime on initial abundance and EE criterion.
(a)~Retained excess $\EEdet$ at the 1~ppmw concentration boundary for
$EE_0=100\%$ and 20\%, for $r=0.04$ (dry solid) and $r=0.10$ (aqueous salt).
(b)~Crossover $\rstar(C_0)$ for $EE_0=100\%$ and EE criteria of 1\%, 10\% and
20\%. The dotted vertical lines mark the baseline $C_0=1000$~ppmw. Neither $r$
value has been measured in an Oxia-relevant matrix.}
\label{fig:regime}
\end{figure}

Because $\rstar$ is independent of $\Dthr$ and the dose profile (Eq.~\eqref{eq:rstar}), the choice of matrix
and amino acid shifts the age boundary but not the regime. The EE criterion, on the
other hand, does matter. For $EE_0=100\%$ and
the 1\% reference, both $r$ values remain destruction-limited
over $C_0=10$--$10^5$~ppmw. At a 20\% criterion, the dry-solid value remains
destruction-limited throughout this range, whereas the aqueous-salt case
becomes EE-limited for $C_0\gtrsim3.1\times10^3$~ppmw
(Fig.~\ref{fig:regime}). Figure~\ref{fig:regime} therefore reports both the
$\EEdet$ curves, which are independent of an assumed EE threshold, and the
EE-threshold-dependent crossover.

\subsection{Two-peak and EE sensitivity from the MOMA-team estimate}
\label{sec:team_sensitivity}

The MOMA-team estimates allow a second, approximate measurement boundary
to be placed on the modeled abundance and EE. For a specified accumulated dose
$D$, the constant-$r$ model gives
\begin{equation}
C_{0,\min}^{\mathrm{peak}}=
\frac{2C_{\mathrm{peak}}\exp(D/\Dthr)}
{1-|EE_0|\exp(-2rD/\Dthr)},
\qquad |EE_0|\exp(-2rD/\Dthr)\gtrsim0.10.
\label{eq:c0peak}
\end{equation}
For $EE_0=1$ and $r>0$, the minor concentration is
$C_0[e^{-x}-e^{-(1+2r)x}]/2$, where $x=D/\Dthr$.
It initially rises as interconversion produces the missing enantiomer, reaches
a maximum at $x=\ln(1+2r)/(2r)$, and then declines through destruction.
A two-peak measurement can therefore occupy a dose interval rather than extend
from zero dose to a single upper limit. The total-abundance crossover in
Eq.~\eqref{eq:rstar} does not classify this additional low-dose constraint.

The first expression in Eq.~\eqref{eq:c0peak} is the initial abundance needed for the \emph{minor}
enantiomer to reach approximately 1~ppmw. It is undefined when the minor
enantiomer is absent, and raising $C_0$ cannot compensate for a case whose EE falls
below the second condition. Unlike Eq.~\eqref{eq:c0min}, the per-peak
requirement depends on $EE_0$ and $r$ as well as on dose.

Table~\ref{tab:team_sensitivity} evaluates this boundary for L-serine with
$C_0=1000$~ppmw, $EE_0=100\%$, the transferred fused-silica+Mg-perchlorate
enhancement, and $r=0.04$. In the baseline radiogenic case (1.57~MGy), a
$10^{7}$-yr surface exposure leaves 0.68~ppmw in the minor peak, just below the
1~ppmw level, although using 1~ppmw as a nominal upper limit on that peak gives
$|EE|>51\%$. In the 100--101~cm layer the same exposure leaves 2.05~ppmw and 71\%
EE, which meets both levels. At $10^{8}$~yr, the 100--101~cm layer retains 46\% EE
but only 0.018~ppmw in the minor peak, while the 199--200~cm layer retains
1.04~ppmw and 67\% EE and meets both levels by a small margin. The 5.6-MGy case lies below the
per-peak level even before cosmic-ray exposure for this initial inventory. If
$r=0.10$ instead, its EE has already fallen to approximately 7\% at that point.
These are modeled sample concentrations before extraction or chromatographic
recovery.

The 199--200~cm result is marginal, as Table~\ref{tab:sens} shows.
Its 4\% minor-peak margin is smaller than the 13\% per-bin Monte Carlo
dose-rate uncertainty and the estimated 5\% endpoint geometry uncertainty;
these central values do not establish a robust detection boundary. It fails if
the radiogenic dose exceeds 1.59~MGy, if the dose rates are more than 3\% higher,
or if destruction is faster by more than 1\%, and it needs an initial abundance
of at least 960~ppmw. The margin is wider when the GCR dose rates or the destruction rate are
halved. The former approximately matches the cited isovaline survival
comparison; the latter is a kinetic sensitivity, not a measured cooling factor. At
100--101~cm, halving the destruction rate or increasing the initial abundance
above $5.5\times10^{4}$~ppmw is sufficient to meet the levels at $10^{8}$~yr,
with the other inputs held fixed. For
$r=0.01$ the minor peak falls below the limit because the EE is 90\%, but the
nominal major-peak bound is then $|EE|>71\%$.

\begin{table}[htbp]
\centering
\caption{Illustrative L-serine comparison with the MOMA-team estimates
of approximately 1~ppmw per enantiomer and EE $\gtrsim10\%$. The model uses
$C_0=1000$~ppmw, $EE_0=100\%$, $r=0.04$, and transferred
fused-silica+Mg-perchlorate kinetics. $C_{\mathrm{minor}}$ and EE are the
predicted surviving sample values; $C_{0,\min}^{\mathrm{peak}}$ is the initial
abundance required for the minor peak alone to reach 1~ppmw at that dose, and
the last column evaluates Eq.~\eqref{eq:eebound} using the modeled major
peak and a nominal minor-peak upper limit of 1~ppmw. These are illustrative
bounds, not confidence limits. The last row includes the radiogenic dose only, so depth is
immaterial.}
\label{tab:team_sensitivity}
\footnotesize
\setlength{\tabcolsep}{4pt}
\begin{tabular}{l l r r r c r}
\toprule
$D_{\mathrm{rad}}$ & GCR exposure, depth & $|EE|$ & $C_{\mathrm{minor}}$ & $C_{0,\min}^{\mathrm{peak}}$ & Both levels & $|EE|$ \\
(MGy) & & (\%) & (ppmw) & (ppmw) & met? & bound (\%) \\
\midrule
1.57 & $10^{7}$~yr, 0--1~cm & 64 & 0.68 & $1.5\times10^{3}$ & no & $>51$ \\
1.57 & $10^{7}$~yr, 100--101~cm & 71 & 2.05 & $4.9\times10^{2}$ & yes & -- \\
1.57 & $10^{8}$~yr, 100--101~cm & 46 & 0.018 & $5.5\times10^{4}$ & no & -- \\
1.57 & $10^{8}$~yr, 199--200~cm & 67 & 1.04 & $9.6\times10^{2}$ & yes & -- \\
5.56 & none, any depth & 35 & $7.4\times10^{-4}$ & $1.4\times10^{6}$ & no & -- \\
\bottomrule
\end{tabular}
\end{table}

\begin{table}[htbp]
\centering
\caption{Sensitivity of the L-serine case of Table~\ref{tab:team_sensitivity} at
$10^{8}$~yr to the main uncertain inputs, changed one at a time from the
baseline ($D_{\mathrm{rad}}=1.57$~MGy, $C_0=1000$~ppmw, $r=0.04$). The other radiogenic doses correspond to K~$=0.05$, 0.20 and 0.354~wt\% (Table~\ref{tab:drad}). The dose-rate
factor scales the GCR dose rates, and the $\Dthr$ factor scales the destruction
dose of serine in the fused-silica+Mg-perchlorate case. The last column states
whether both MOMA-team levels are met at 199--200~cm and 100--101~cm.}
\label{tab:sens}
\scriptsize
\setlength{\tabcolsep}{3pt}
\begin{tabular}{l r r r r c}
\toprule
 & \multicolumn{2}{c}{100--101~cm} & \multicolumn{2}{c}{199--200~cm} & Levels met \\
\cmidrule(lr){2-3}\cmidrule(lr){4-5}
Change from baseline & $C_{\mathrm{minor}}$ (ppmw) & $|EE|$ (\%) & $C_{\mathrm{minor}}$ (ppmw) & $|EE|$ (\%) & 2~m, 1~m \\
\midrule
Baseline & 0.018 & 46 & 1.04 & 67 & yes, no \\
$D_{\mathrm{rad}}=0.78$~MGy & 0.098 & 54 & 4.46 & 77 & yes, no \\
$D_{\mathrm{rad}}=3.13$~MGy & $5.7\times10^{-4}$ & 35 & 0.041 & 50 & no, no \\
$D_{\mathrm{rad}}=5.56$~MGy & $2.3\times10^{-6}$ & 22 & $1.9\times10^{-4}$ & 32 & no, no \\
Dose rates $\times0.5$ & 0.27 & 59 & 1.87 & 71 & yes, no \\
Dose rates $\times1.5$ & $1.1\times10^{-3}$ & 37 & 0.57 & 63 & no, no \\
$\Dthr\times0.7$ & $3.7\times10^{-4}$ & 33 & 0.16 & 56 & no, no \\
$\Dthr\times2$ & 1.31 & 68 & 7.26 & 82 & yes, yes \\
$r=0.01$ & $5.9\times10^{-3}$ & 83 & 0.30 & 90 & no, no \\
$r=0.10$ & 0.029 & 15 & 1.99 & 36 & yes, no \\
$r=0.20$ & 0.033 & 2 & 2.71 & 13 & yes, no \\
$C_0=100$~ppmw & $1.8\times10^{-3}$ & 46 & 0.10 & 67 & no, no \\
$C_0=10^{4}$~ppmw & 0.18 & 46 & 10.4 & 67 & yes, no \\
$C_0=10^{5}$~ppmw & 1.8 & 46 & 104 & 67 & yes, yes \\
\bottomrule
\end{tabular}
\end{table}

\subsection{Minimum initial abundance at a specified depth and age}
\label{sec:c0min}

The preceding sections give a maximum exposure age for an assumed $C_0$.
Inverting the relation gives the initial abundance required at a specified depth
and age. Setting $C(z,t_{\mathrm{exp}},t_{\mathrm{dep}}) = \Cthr$ in
Eq.~\eqref{eq:destr} gives
\begin{equation}
C_{0,\min}(z, t_{\mathrm{exp}}, t_{\mathrm{dep}}) = \Cthr \,
\exp\!\left[ D_{\mathrm{total}}(z,t_{\mathrm{exp}},t_{\mathrm{dep}})/\Dthr \right],
\label{eq:c0min}
\end{equation}
the smallest initial concentration required for a given amino acid to meet the
1~ppmw R5 reference concentration in the delivered bulk sample before
extraction and analytical recovery. Unlike the regime
criterion of Eq.~\eqref{eq:rstar}, this quantity depends on the dose field and on $\Dthr$.
With the total dose of Eq.~\eqref{eq:dtotal} the requirement factorizes,
\begin{equation}
C_{0,\min}(z, t_{\mathrm{exp}}, t_{\mathrm{dep}}) =
 \Cthr\,\underbrace{\exp\!\left[D_{\mathrm{rad}}(t_{\mathrm{dep}})/\Dthr\right]}_{\text{depth-independent factor}}
 \;\underbrace{\exp\!\left[\dot D_{\mathrm{GCR}}(z)\,t_{\mathrm{exp}}/\Dthr\right]}_{\text{reduced by drilling}},
\label{eq:c0minfactor}
\end{equation}
into a depth-independent factor set by the deposition age and a factor that
depends on depth and exposure age.

Because Eq.~\eqref{eq:c0min} is exponential in dose, a modest reduction of the
second factor at depth can produce a much larger reduction in $C_{0,\min}$,
while the first factor is invariant. The depth-independent factor alone is
$1.4\times$ for L-valine and $2.2\times$ for L-aspartic acid, rising to
$18\times$ for L-glutamic acid, $39\times$ for L-serine, $2.3\times10^{2}$ for
L-isovaline and $6.1\times10^{2}$ for L-alanine in the fused-silica+Mg-perchlorate
case. Equivalently, it sets a minimum initial abundance below which the amino acid
cannot reach the reference concentration at any depth or exposure age: 610~ppmw
for L-alanine, 230~ppmw for L-isovaline and 39~ppmw for L-serine.

\begin{figure}[htbp]
\centering
\includegraphics[width=0.9\textwidth]{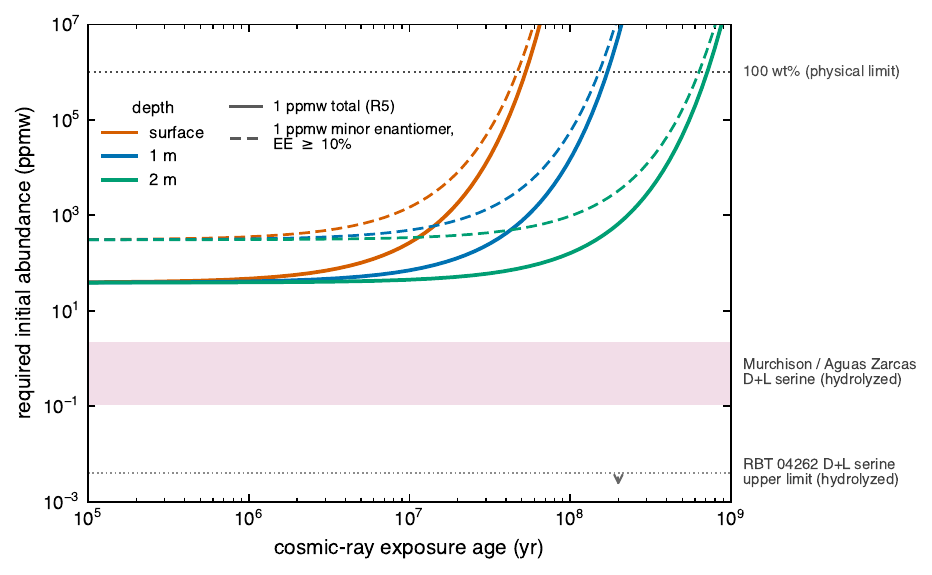}
\caption{Initial abundance of L-serine required to reach 1~ppmw total (solid) and
1~ppmw in the minor enantiomer with $|EE|\geq10\%$ (dashed), as a function of
cosmic-ray exposure age at three depths, for the fused-silica+Mg-perchlorate case,
$r=0.04$ and $D_{\mathrm{rad}}=1.57$~MGy. The total-analyte curves approach 39~ppmw at zero cosmic-ray exposure;
the per-peak curves instead approach about 310~ppmw. Dashed curves are shown
only where the EE criterion is met. The comparison band shows measured
D+L serine in acid-hydrolyzed Murchison and Aguas Zarcas extracts
(0.105--2.25~ppmw, from Table~1 of \citealp{Glavin2021}); the downward arrow
marks the combined serine upper limit in hydrolyzed RBT~04262
($<0.004$~ppmw; \citealp{Callahan2013}). These present-day extractable abundances
are contextual comparisons, not estimates or upper bounds on the initial
Oxia inventory. Hydrolysis and MOMA derivatization recover different molecular
pools, and meteoritic proteinogenic amino acids may include contamination.}
\label{fig:required}
\end{figure}

At a $10^{7}$~yr constant-depth exposure the inferred $C_{0,\min}$ ranges from
1.7~ppmw for L-valine to $1.8\times10^{4}$~ppmw for L-alanine at the surface,
and from 1.5 to $1.7\times10^{3}$~ppmw in the 100--101~cm layer. At
$10^{8}$~yr, comparable to measured exposure ages of recently exhumed martian
mudstone~\citep{Farley2014}, four of the six amino acids require
$C_0>10^6$~ppmw ($>100$~wt\%) at the surface, exceeding the total sample mass.
In the 100--101~cm layer the four least radiation-sensitive amino acids span
2.4--$1.5\times10^{4}$~ppmw, while L-isovaline and L-alanine have no physically
allowed solution. Figure~\ref{fig:required} shows the requirement for serine as a
function of exposure age and depth, together with abundances measured in natural
samples. Greater
shielding lowers the inferred initial abundance by orders of magnitude in
this model, but not below the level set by the radiogenic dose. Chiral measurement
additionally depends on analytical recovery and EE precision.

\begin{table}[htbp]
\centering
\caption{Minimum initial abundance $C_{0,\min}$ (ppmw) required to retain
1~ppmw after GCR exposure and the 1.57-MGy radiogenic dose accumulated since
3.9~Ga. The fused-silica+Mg-perchlorate kinetics are transferred across
amino acids; the last column is the depth-independent factor
$\exp(D_{\mathrm{rad}}/\Dthr)$. $\dagger$ denotes $>10^{6}$~ppmw ($>100$~wt\%),
which exceeds the sample mass. Surface values omit solar energetic particles
and are lower limits. Layer depths are in cm. Values use the boundary-corrected profile and are
rounded to two significant figures.}
\label{tab:c0min}
\scriptsize
\setlength{\tabcolsep}{4pt}
\begin{tabular}{l r r r r r r r}
\toprule
 & \multicolumn{3}{c}{$t_{\mathrm{exp}} = 10^{7}$~yr} & \multicolumn{3}{c}{$t_{\mathrm{exp}} = 10^{8}$~yr} & Factor \\
\cmidrule(lr){2-4}\cmidrule(lr){5-7}
Amino acid & 0--1 & 100--101 & 199--200 & 0--1 & 100--101 & 199--200 & $e^{D_{\mathrm{rad}}/\Dthr}$ \\
\midrule
L-valine        & 1.7 & 1.5 & 1.4 & 7.8 & 2.4 & 1.6 & $1.4\times$ \\
L-aspartic acid & 3.3 & 2.5 & 2.3 & $1.4{\times}10^{2}$ & 7.9 & 3.0 & $2.2\times$ \\
L-glutamic acid & 81  & 29  & 20 & $\dagger$ & $1.9{\times}10^{3}$ & 55 & $18\times$ \\
L-serine        & $2.6{\times}10^{2}$ & 70 & 45 & $\dagger$ & $1.5{\times}10^{4}$ & $1.6{\times}10^{2}$ & $39\times$ \\
L-isovaline     & $3.9{\times}10^{3}$ & $5.5{\times}10^{2}$ & $2.8{\times}10^{2}$ & $\dagger$ & $\dagger$ & $1.9{\times}10^{3}$ & $2.3{\times}10^{2}$ \\
L-alanine       & $1.8{\times}10^{4}$ & $1.7{\times}10^{3}$ & $7.8{\times}10^{2}$ & $\dagger$ & $\dagger$ & $7.4{\times}10^{3}$ & $6.1{\times}10^{2}$ \\
\bottomrule
\end{tabular}
\end{table}

For the regional 5.6-MGy $\beta+\gamma$ case, the same calculation gives
Table~\ref{tab:grs}. The large factors for radiation-sensitive amino acids follow from the
exponential $D_{\mathrm{rad}}/\Dthr$; they do not establish that this laboratory
matrix enhancement applies at Oxia. At $10^{8}$~yr and 100--101~cm, only
valine and aspartic acid remain below the physical $10^{6}$~ppmw bulk limit.

\begin{table}[htbp]
\centering
\caption{Minimum initial abundance for the regional GRS $\beta+\gamma$ radiogenic dose at a
3.9-Ga deposition age, $D_{\mathrm{rad}}=5.6$~MGy, with the same transferred
fused-silica+Mg-perchlorate kinetics and 1~ppmw total-analyte reference as
Table~\ref{tab:c0min}. The radiogenic factor is $\exp(D_{\mathrm{rad}}/\Dthr)$,
independent of depth and cosmic-ray exposure. The last column includes
$10^{8}$~yr of cosmic-ray exposure at 100--101~cm. Values above
$10^{6}$~ppmw exceed the entire sample mass.}
\label{tab:grs}
\begin{tabular}{l r r}
\toprule
Amino acid & Radiogenic factor & $C_{0,\min}$ at 100--101~cm (ppmw) \\
\midrule
L-valine & $3.2\times$ & 5.5 \\
L-aspartic acid & $1.6\times10^{1}$ & $6.0\times10^{1}$ \\
L-glutamic acid & $2.8\times10^{4}$ & $3.0\times10^{6}$ \\
L-serine & $4.4\times10^{5}$ & $1.7\times10^{8}$ \\
L-isovaline & $2.4\times10^{8}$ & $1.6\times10^{12}$ \\
L-alanine & $7.8\times10^{9}$ & $2.7\times10^{14}$ \\
\bottomrule
\end{tabular}
\end{table}

\subsection{Radiogenic bound on the benefit of drilling}
\label{sec:floor}

\begin{figure}[htbp]
\centering
\includegraphics[width=\textwidth]{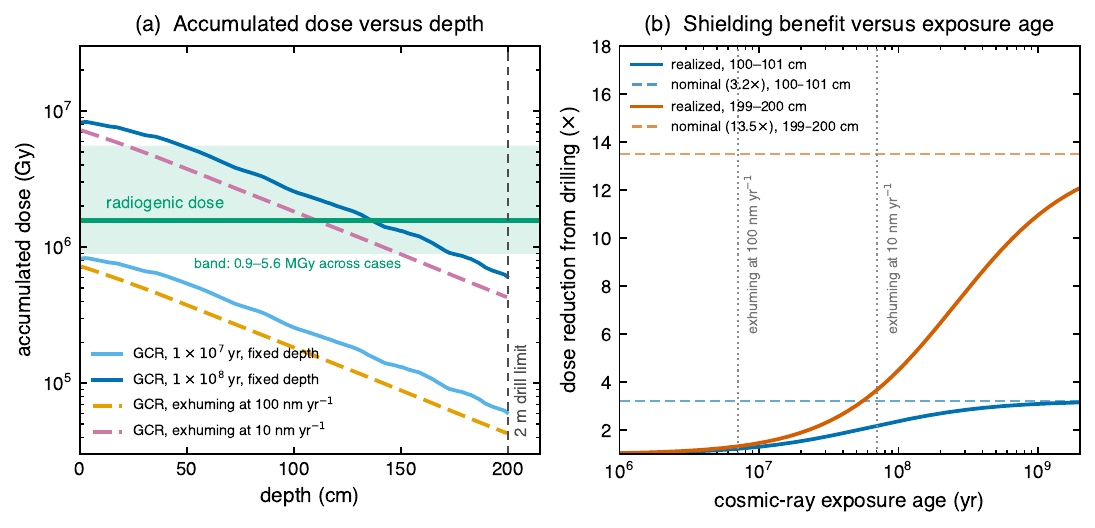}
\caption{Radiogenic dose and the benefit of drilling. (a) GCR dose versus
depth for constant-depth and steadily exhuming horizons, compared with the
depth-independent radiogenic dose. Shading spans the radiogenic dose of the baseline composition at 3~Ga
(0.9~MGy) and that of the regional GRS abundances at 3.9~Ga (5.6~MGy). Where a
curve crosses the radiogenic level, the two doses are equal. (b) Total-dose
reduction at 100--101 and 199--200~cm versus exposure age; dashed lines are
the GCR-only shielding factors. Dotted lines mark effective ages for 10 and
100~nm\,yr$^{-1}$ exhumation.}
\label{fig:floor}
\end{figure}

Sampling depth reduces the cosmic-ray dose but leaves the radiogenic dose
unchanged, so the reduction
achieved by drilling is the ratio of \emph{total} doses,
\begin{equation}
\frac{D(0, t_{\mathrm{exp}})}{D(z, t_{\mathrm{exp}})} =
\frac{\dot D_{\mathrm{GCR}}(0)\,t_{\mathrm{exp}} + D_{\mathrm{rad}}}
     {\dot D_{\mathrm{GCR}}(z)\,t_{\mathrm{exp}} + D_{\mathrm{rad}}},
\label{eq:realized}
\end{equation}
which tends to the nominal shielding factor only when the cosmic-ray term
dominates at both depths, and to unity when it dominates at neither
(Fig.~\ref{fig:floor}).

For the modeled profile the nominal factors are 3.2 in the 100--101~cm layer
and 13.5 in the 199--200~cm layer. The realized factors are 1.04 and 1.05 at $t_{\mathrm{exp}}=10^{6}$~yr,
1.3 and 1.5 at $10^{7}$~yr, 2.4 and 4.5 at $10^{8}$~yr, and 3.1 and 11 at
$10^{9}$~yr. At $10^{9}$~yr the total-dose reduction at 2~m is still below the GCR-only
limit of 13.5; that limit is approached asymptotically as GCR dose dominates.
For an actively exhuming surface the effective exposure age is
$\lambda/E$ (Section~\ref{sec:exhumation}), $7\times10^{6}$~yr at
$100$~nm\,yr$^{-1}$, and the available reduction is then less than a factor of
two at any accessible depth.

\subsection{Effect of an earlier aqueous period on the initial excess}

\begin{figure}[htbp]
\centering
\includegraphics[width=\textwidth]{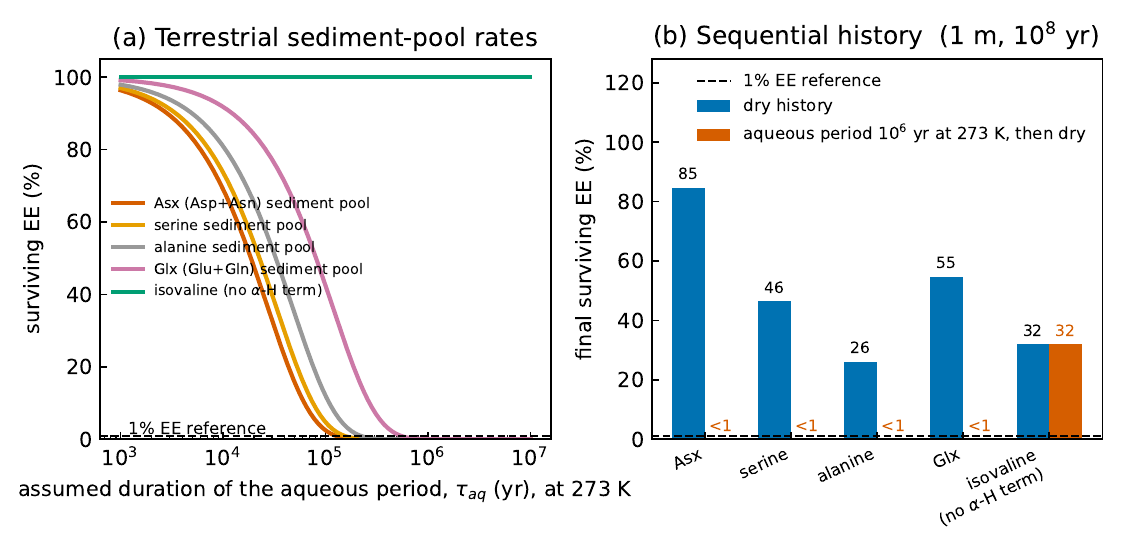}
\caption{Effect of an earlier aqueous period on EE, using terrestrial
sediment-pool rates from \citet{Steen2013}, as corrected by
\citet{SteenCorrection2015}.
(a)~EE versus the assumed duration of the aqueous period at 273~K for Asx (Asp+Asn), Glx
(Glu+Gln), serine, and alanine pools; the isovaline curve sets only the
$\alpha$-hydrogen-mediated term to zero. The horizontal line is the
1\% EE reference. (b)~Sequential history at 1~m and $10^{8}$~yr constant-depth cosmic-ray exposure, with the 3.9-Ga radiogenic dose added in both cases,
and with or without a $10^{6}$~yr aqueous period in sediment; bar labels give the
final EE in percent.}
\label{fig:aqueous}
\end{figure}

The corrected terrestrial-sediment rates illustrate how much an
earlier aqueous period can control the EE inherited by the radiation epoch
(Fig.~\ref{fig:aqueous}). At 273~K the Asx sediment pool falls below the 1\%
reference after ${\sim}10^{5}$~yr and Glx after ${\sim}5\times10^{5}$~yr, with
serine and alanine between; in a sequential history (1~m, $10^{8}$~yr exposure) a
$10^{6}$~yr aqueous period drives all four proteinogenic pools below 1\%. For isovaline the aqueous $\alpha$-hydrogen term is zero, and its
radiation term uses the same $r$ as the other amino acids. Pre-radiation
history can therefore determine the EE entering the radiation epoch. These are
terrestrial-sediment rates and give only order-of-magnitude timescales;
pure-compound and Oxia-specific kinetics are needed for firmer estimates.
Setting the $\alpha$-hydrogen term to zero for isovaline omits only that
pathway and does not imply immunity to other changes during aqueous alteration.

\section{Discussion}
\label{sec:discussion}

\subsection{Initial abundance and the 1~ppmw survival boundary}
\label{sec:abundance}

In the homogeneous first-order model with $C_0=1000$~ppmw and
$EE_0=100\%$, abundance reaches 1~ppmw before EE falls to the 1\%, 10\%,
or 20\% reference for either baseline $r$. MOMA's EE precision remains an
independent measurement requirement. The initial inventory
at Oxia Planum is unknown. A compound-specific comparison is more informative
than comparing serine with total amino acids or organic carbon.
\citet{Callahan2013} report $<2$~ppb for each serine enantiomer in the
hydrolyzed RBT~04262 extract. The combined D+L serine abundances in the
hydrolyzed Murchison and Aguas Zarcas extracts of \citet{Glavin2021}
span approximately 0.105--2.25~ppmw. These measurements are well below the
modeled initial serine requirements, but sample processing, terrestrial
contamination, and subsequent preservation prevent a direct reconstruction
of the original martian inventory from them.

The 201--273~$\mu$g~C\,g$^{-1}$ reported for Gale mudstone by
\citet{Stern2022} refers to refractory organic carbon inferred from
high-temperature combustion, not to a serine concentration or an upper bound
on the initial organic inventory. Serine contains 34.3\% carbon by mass, so
340--960~ppmw serine would contain approximately 117--329~ppmw carbon.
Comparing the two quantities without this conversion overstates the abundance
constraint; even with it, different sites, ages and molecular pools prevent a
like-for-like inference. The $C_0=1000$~ppmw baseline is an illustrative enriched
serine population, not an abundance estimated for Oxia.

The inversion in Eq.~\eqref{eq:c0min} expresses that uncertainty as a required
initial abundance for each depth and exposure age. Extraction losses would
raise the requirement further. Existing SAM detections of
chlorohydrocarbons, thiophenic and aromatic compounds, alkanes, and products of tetramethylammonium hydroxide (TMAH)
thermochemolysis~\citep{Freissinet2015,Eigenbrode2018,Freissinet2025,Williams2026}
do not measure EE or constrain an initial amino acid inventory, so they cannot
test the predicted ordering of abundance and EE loss.

\subsection{Drill depth, exposure age, and target selection at Oxia Planum}
\label{sec:drill}

Increasing depth reduces the cosmic-ray part of the accumulated dose and lowers
$C_{0,\min}$ exponentially, but not below the limit set by the depth-independent radiogenic dose
(Section~\ref{sec:floor}). An exhuming sample follows a
time-dependent $z(t)$. An erosion rate of ${\sim}100$~nm\,yr$^{-1}$ would move a
horizon by ${\sim}10$~m in $10^{8}$~yr, well beyond the 2-m constant-depth slab;
the same rate corresponds to an effective exposure age of only
${\sim}7\times10^{6}$~yr in Eq.~\eqref{eq:exhum}, at which the radiogenic term
dominates the total dose at every accessible depth. Targets should therefore be deep and geomorphically
recently exposed, although for a rapidly exhuming surface the exposure age is
already short and the reduction in total dose from depth is
correspondingly small. The survival gain can nevertheless be appreciable
because abundance depends exponentially on dose.
Regional geomorphology indicates
ongoing exhumation and crater obliteration at Oxia~\citep{QuantinNataf2021}, but
does not assign a ${\lesssim}100$~Myr exposure age to a specific phyllosilicate
outcrop. The ${\sim}78$~Myr cosmogenic age at Gale~\citep{Farley2014} and erosion
estimates for martian light-toned sedimentary rocks~\citep{KiteMayer2017} provide
context but no exposure age for Oxia. The ${\sim}3$--4-m resistant low-albedo unit is
interpreted as an airfall or paleo-ash deposit preferentially cemented by
groundwater in paleotopographic lows and is discontinuous
~\citep{Harris2024}. Where present, it may have shielded underlying material;
the accumulated protection depends on the burial and exhumation history.

Recent orbital studies show mineralogical variation across Oxia Planum.
\citet{McNeil2025} mapped ${\sim}85$~km$^2$ of hydrated silica, predominantly
opal-A, near the clay-plain/fan transition and in southern topographic lows,
including candidate rover-accessible deposits. Authigenic, groundwater,
pedogenic, and detrital origins remain under consideration; consequences for
organic preservation have not been established. \citet{TorresAure2026} links Oxia's basal Mg-rich
smectite/ferroan-saponite/vermiculite unit regionally to Mawrth Vallis and
identifies an overlying Fe$^{3+}$-rich nontronite unit and intervening cratered
paleosurface elsewhere in the regional stratigraphy. The
fused-silica+Mg-perchlorate laboratory factor used here cannot be assigned to
natural opal or to these clay units. Target assessment therefore requires mineralogy, stratigraphic
position, former capping, local aqueous history, depth, and exposure age.
For a $10^{8}$~yr constant-depth exposure in this high-destruction case, the
required initial abundance exceeds $10^{3}$~ppmw for L-glutamic acid, L-serine,
L-isovaline and L-alanine even at 100--101~cm (Table~\ref{tab:c0min}), before
extraction losses.

\subsection{Mineral-matrix dependence of the radiogenic channel}
\label{sec:reconcile}

Published preservation estimates use different destruction kinetics.
\citet{KminekBada2006} computed the radiogenic dose and concluded that
an early-biosphere amino acid signature is not limited by radiolytic
decomposition provided the material is adequately shielded from space radiation.
\citet{Pavlov2022} measured destruction rates in Mars-analog mineral matrices
an order of magnitude or more faster than in pure solids and concluded that the
upper two meters are unfavorable for the preservation of ancient amino acids. These conclusions depend
on both the radiogenic dose and the destruction constants, and the comparison
below varies each input in turn.

Table~\ref{tab:floor} gives the depth-independent factor
$\exp(D_{\mathrm{rad}}/\Dthr)$ across the matrix cases. At 1.57~MGy, pure-solid constants raise the required initial
abundance by at most about 20\%, consistent with the Kminek--Bada conclusion.
The fused-silica+Mg-perchlorate enhancement instead raises it by up to
$6.1\times10^{2}$. At 5.6~MGy the same transferred enhancement gives a factor of
${\sim}10^{10}$ for alanine and ${\sim}5\times10^{5}$ for serine
(Table~\ref{tab:grs}). For $C_0=1000$~ppmw, the radiogenic dose alone
then drives serine, isovaline and alanine below the 1~ppmw reference
at every depth. This exclusion applies to the stated initial abundance and
matrix case, not to all possible initial inventories. At 1.57~MGy the intermediate multi-amino acid mixture case leaves the factor
below 1.5 for every amino acid; at 5.6~MGy that statement no longer holds.

Which case applies at Oxia Planum is not established. The enhancement factors
were derived from irradiations in fused silica and dry magnesium perchlorate, not
in Fe/Mg phyllosilicate or opal-A, and the transfer across both amino acids and
matrix is an assumption (Section~\ref{sec:destruction}). Depending on this factor, the
depth-independent channel is either negligible or dominant. A matrix-resolved
determination of $\Dthr$ in candidate Oxia mineralogies is therefore the most
useful measurement for this problem, more so than any further refinement of the
transport calculation.

\begin{table}[htbp]
\centering
\caption{Depth-independent factor $\exp(D_{\mathrm{rad}}/\Dthr)$ by which the
radiogenic dose accumulated since deposition raises the required initial
abundance, for $D_{\mathrm{rad}}=1.57$~MGy at a 3.9-Ga deposition age. The
factor applies at every depth and drilling does not reduce it. Matrix enhancement
factors are $\times2.1$ (multi-amino acid mixture), $\times21$ (fused silica)
and $\times35$ (fused silica$+$Mg-perchlorate), as in Table~\ref{tab:kinetics}. With pure-solid constants the channel is a small
correction; with the laboratory mineral enhancement it dominates for the more
radiation-sensitive amino acids.}
\label{tab:floor}
\footnotesize
\setlength{\tabcolsep}{9pt}
\begin{tabular}{l r r r r}
\toprule
Amino acid & pure solid & mixture & $+$silica & $+$silica$+$Mg(ClO$_4)_2$ \\
\midrule
L-valine        & 1.0 & 1.0 & 1.2  & 1.4   \\
L-aspartic acid & 1.0 & 1.0 & 1.6  & 2.2   \\
L-glutamic acid & 1.1 & 1.2 & 5.6  & 18    \\
L-serine        & 1.1 & 1.2 & 9.0  & 39    \\
L-isovaline     & 1.2 & 1.4 & 26   & $2.3\times10^{2}$ \\
L-alanine       & 1.2 & 1.5 & 47   & $6.1\times10^{2}$ \\
\bottomrule
\end{tabular}
\end{table}

\subsection{Kinetic-dataset dependence of the ordering of amino acids}
\label{sec:aspartic}

The baseline $C_{0,\min}$ values derive from a single kinetic compilation, so
the ranking of amino acids by radiation resistance is uncertain.
For example, the converted DSC values of \citet{IglesiasGroth2011} give
$D_{37}^{\mathrm{Asp}}=69.5$~MGy and $D_{37}^{\mathrm{Ala}}=8.55$~MGy, whereas
fits to the multi-dose HPLC concentrations of \citet{KminekBada2006} give
6.25 and 8.87~MGy, respectively. The ordering of aspartic acid and alanine
therefore reverses between
published datasets. Differences in irradiation protocol, observable, fitting,
and sample form may explain part of the discrepancy, and the available data do
not show which set applies to Oxia Planum. The disagreement leaves the algebraic form of the abundance inversion
unchanged, but changes its numerical values, exposure-age limits and the
ranking of amino acids. Resolving it requires a common-protocol, multi-dose
determination of destruction and true L$\leftrightarrow$D interconversion for
the same amino acids in candidate Oxia mineral--oxychlorine mixtures at low
dose rate.

The same kinetic data bear on one possible observational test. If a co-located
suite of proteinogenic amino acids shares one initial EE magnitude, one
accumulated dose, one $r$ and one matrix scaling, Eq.~\eqref{eq:eedecay} gives
\begin{equation}
\ln\!\left(1/|EE_i|\right) = \ln\!\left(1/|EE_{0}|\right)
+ 2rD\,\frac{1}{\Dthr^{(i)}},
\label{eq:crossspecies}
\end{equation}
so that $\ln(1/|EE_i|)$ falls on a straight line in $1/\Dthr^{(i)}$, with
initial concentrations dropping out. The slope identifies $2rD$ (or $2rDF$ for an
unknown common matrix enhancement $F$). A measured suite could therefore be
checked against this line, although the test uses $|EE|$ only and cannot check
for a common sign. Species-dependent $r_i$, different initial excesses or
matrices, or a mixture of labile and refractory domains would break the
relation, and the dataset disagreement above shows that the $\Dthr$ values on the
abscissa are themselves uncertain.

\subsection{Abiotic excesses and the diagnostic value of a detection}
\label{sec:abiotic}

Interpretation of a detected EE is amino acid-specific. For a freshly
synthesized organic pool, circularly polarized irradiation of
extraterrestrial-ice analogs can produce small primary amino acid excesses, up
to a few percent in the cited experiments~\citep{Modica2014, Glavin2020}.
Enantioselective adsorption or catalysis on chiral mineral surfaces provides a distinct possible source or local redistribution mechanism. Subsequent
chemistry or phase partitioning may increase a pre-existing excess; these
processes are not included in the radiation model.
Parent-body aqueous processing can amplify or redistribute such excesses.
Large L-excesses reported for proteinogenic amino acids in some carbonaceous
meteorites cannot be separated from terrestrial contamination by EE
alone; isotopic composition, blanks, and petrographic context are needed to
establish an indigenous component~\citep{Glavin2021}. That study also reports
isotopic evidence for a non-terrestrial contribution to L-glutamic-acid excess
in Murchison, so a proteinogenic identity does not exclude abiotic excess.
Isovaline provides a
stronger case for parent-body processing because it is non-proteinogenic and its
indigenous L-excess covaries with alteration state. CI and CM meteorites show
L-isovaline enrichment associated with aqueous alteration
~\citep{GlavinDworkin2009, Glavin2021}. The magnitude and sign of an indigenous
excess depend on molecular identity and alteration history. Nonlinear reaction networks with autocatalysis and
enantiomeric cross-inhibition can amplify a small imbalance
~\citep{Brandenburg2005}, although no such network has been established for Oxia
materials. EE magnitude alone cannot distinguish these abiotic sources from
biology.

The radiation calculation takes $EE_0$ as an initial condition and evolves it
only through non-enantioselective destruction and symmetric
L$\leftrightarrow$D interconversion (Section~\ref{sec:racchannel}). The origin
of $EE_0$ and any amplification before irradiation therefore lie outside the
model, as do heterogeneous domains, chiral mineral surfaces, enantioselective
adsorption, and differential analytical recovery. The asymmetry in dissociative
electron attachment measured for bromocamphor by \citet{DreilingGay2014} cannot
be transferred quantitatively to amino acid destruction. Astrobiological
interpretation therefore requires paired-enantiomer measurements across several
amino acids, molecular distributions across the wider organic inventory,
blanks, spatial controls, contamination assessment, and mineralogical and
stratigraphic context~\citep{Goesmann2017,Vago2017}.

\subsection{Limitations}
\label{sec:limits}

The main limitation on the chirality calculation is the ratio $r$ between true
radioracemization and radiolytic destruction (Section~\ref{sec:racchannel}); the
depth--abundance inversion is independent of $r$. Available measurements use
high-dose-rate $\gamma$-irradiation of order kGy\,h$^{-1}$
~\citep{Bonner1979, Pavlov2022}, roughly eight to ten orders of magnitude above
the modeled martian GCR dose rate, and provide no compound- and matrix-specific
value for Oxia. \citet{Pavlov2022} used 1.8~Gy\,s$^{-1}$ and argue that the
fraction of molecules in excited states at any moment, about $10^{-14}$, is too
small for them to interact, so the high rate should not alter the degradation.
The excited-state estimate suggests that the high laboratory dose rate may have
little effect on degradation, although dose-rate dependence in solid amino acids
has not, to our knowledge, been tested experimentally. A common rescaling of
both channels leaves the regime criterion unchanged. Differential dose-rate effects, however, could shift $r$ in either
direction. Increasing $r$ by a factor of 8.3 in the dry-solid case or 3.3
in the aqueous-salt case would cross the baseline boundary.

The exponential extrapolation also assumes concentration-independent kinetics.
The mineral enhancement was measured for dilute laboratory mixtures; very large
inferred initial abundances could change radical scavenging, molecular contacts
and energy partition. Values approaching bulk mass fractions should therefore
be read as extrapolations beyond the tested concentration regime, even when
they remain below the formal $10^{6}$~ppmw mass limit.

The model assumes equal destruction rates for the two enantiomers, and
paired-enantiomer irradiations are needed to test that assumption. The
proteinogenic $\Dthr$ values come from \citet{IglesiasGroth2011}, with the
serine value checked against one independent survival measurement.

For application to MOMA, the model stops at the preserved bulk concentration in
the delivered subsample. Mapping these quantities to measured EE additionally requires one-pot extraction
and derivatization, transfer and trapping, chromatographic resolution, peak
identification and integration, and characterization of the relative end-to-end
response of the two enantiomers~\citep{Boulesteix2023,Guzman2020}. A
test across amino acids must use compounds resolved within the same DMF-DMA
subsample; an earlier subsample derivatized with MTBSTFA ($N$-tert-butyldimethylsilyl-$N$-methyltrifluoroacetamide) cannot establish identical composition.
Blanks, carryover, and terrestrial contamination also require control
~\citep{Goesmann2017}. Finally, the dose profile is a single hydrated-clay scenario, checked against
a silicon-detector RAD reference at the surface and compared with the independent calculation of \citet{Pavlov2022}
at depth, with fixed solar modulation and atmospheric column density. Uncertainties in composition, solar cycle, and burial and exhumation history
affect the absolute exposure ages.

The radiogenic channel has its own uncertainties. Its amplitude depends on the
potassium content of the host rock, which has not been measured at Oxia Planum,
and spans 0.8--5.6~MGy at 3.9~Ga for K~$=0.05$~wt\% to the regional GRS value
(Table~\ref{tab:drad}). The 2~m serine result of Table~\ref{tab:sens} fails above
1.6~MGy, which corresponds to K~${\sim}0.10$~wt\% at 3.9~Ga. This spread
reflects the unknown composition and not a statistical uncertainty interval.
The $\gamma$-ray abundances are themselves regional surface averages over a
footprint far larger than an outcrop. The infinite-matrix
treatment assumes spatial averaging of sources and absorbers; organic molecules concentrated in domains distant from the
radionuclide-bearing phases would receive less, and organic molecules adsorbed onto such
phases more. This channel is delivered at about $10^{-4}$~Gy\,yr$^{-1}$, and its $\beta$ and $\gamma$ radiation quality
is closer to that of the $^{60}$Co irradiations from which $\Dthr$ is derived.
Whether it matters depends on the matrix enhancement
(Section~\ref{sec:reconcile}).

\section{Conclusions}

Radiation limits on amino acid chirality depend on total survival and on the
concentrations of both enantiomers. The analytic crossover
$r^{*}=\ln(|EE_0|/\EEthr)/[2\ln(C_0/\Cthr)]$ specifies which preservation
criterion fails first. At fixed $r$ it is independent of the characteristic
destruction dose and the dose profile. For $C_0=1000$~ppmw, $EE_0=100\%$,
and $r=0.04$ or 0.10, the total abundance reaches 1~ppmw while 58\% or 25\% EE
remains. Earlier experiments established the slower racemization of surviving
amino acids; the result here quantifies the dependence of that ordering on the
initial population and the selected measurement criteria. It does not establish
that two enantiomer peaks are measurable at 1~ppmw total abundance.

For serine with the transferred fused-silica--perchlorate kinetics and a
1.57-MGy radiogenic dose, the provisional 1~ppmw-per-enantiomer and 10\% EE
criteria require initial abundances of 340 and 960~ppmw at 2~m after
$10^{7}$ and $10^{8}$~yr of cosmic-ray exposure. At 1~m the corresponding
requirements are 490 and $5.5\times10^{4}$~ppmw. The $10^{8}$-yr, 2-m case
with $C_0=1000$~ppmw is marginal: a radiogenic dose above about 1.59~MGy or
a roughly 1\% increase in destruction rate removes the two-peak solution.
The required initial concentrations exceed the serine abundances in the
meteorite samples considered, but those present-day measurements do not bound
the initial Oxia inventory. Compound-specific recovery and EE precision remain
necessary to convert these sample-level criteria into instrument predictions.

In the closed-host model, organics retained since deposition 3.9~Ga ago
accumulate 1.57~MGy for K~$=0.10$~wt\% or 5.6~MGy for the regional GRS
composition. This buried radiogenic contribution is largely insensitive to
sampling depth. In the lower-dose case it reduces the surface-to-2-m total-dose
ratio to 1.5 after $10^{7}$~yr, compared with a GCR-only ratio of 13.5.
The effect on molecular survival depends strongly on the mineral-dependent
destruction rate. Younger organic residence times, radionuclide redistribution,
and different host matrices would change the inferred abundance floor.

A non-detection would constrain the combination of initial abundance and
preservation history. A detected excess would
likewise require assessment of abiotic production, contamination and analytical
bias. The most useful next measurements are paired destruction and
interconversion rates in candidate Oxia mineral mixtures, enantiomer-specific
analytical recovery, and constraints on the host composition and organic
residence history. These would determine whether the conditional boundaries
derived here describe the samples available to the ExoMars drill.

\section*{Declaration of competing interest}

The author declares that there are no known competing financial interests or
personal relationships that could have appeared to influence the work reported
in this paper.

\section*{Data and code availability}

The data and code supporting this study are available upon reasonable request to
the author. The Mars Odyssey Gamma Ray Spectrometer element-concentration maps
are archived in the NASA Planetary Data System~\citep{GRSdata,Boynton2007}, and
the Geant4 toolkit is distributed by the Geant4 collaboration at CERN
(\url{https://geant4.web.cern.ch};
\citealp{Agostinelli2003,Allison2016}).

\section*{Acknowledgements}

Funding for this research was provided by the NYU Abu Dhabi Research Institute
grant CG014. Computational facilities were provided by the NYUAD High
Performance Computing (HPC) facility and the Center for Astrophysics and Space
Science (CASS). We thank the Geant4 collaboration at CERN for the simulation
toolkit. We thank Axel Brandenburg for helpful comments, and Guillaume
Leseigneur and Sandra Siljestr\"om of the MOMA team for their feedback.

\end{document}